\documentclass[onecolumn,fleqn,usenatbib]{mnras}

\usepackage{newtxtext,newtxmath}

\usepackage[T1]{fontenc}

\DeclareRobustCommand{\VAN}[3]{#2}
\let\VANthebibliography\thebibliography
\def\thebibliography{\DeclareRobustCommand{\VAN}[3]{##3}\VANthebibliography}

\usepackage{graphicx}	
\usepackage{amsmath}	
\usepackage{multicol}
\usepackage{bm}
\usepackage{cancel}
\usepackage[dvipsnames]{xcolor}

\newcommand{\f}{\frac}
\newcommand{\dd}{\mathrm{d}}

\newcommand{\ii}{\mathrm{i}}

\title[Laplace Surface]{The Laplace Surface of a Circumplanetary Disc }

\author[Lubow \& Ogilvie ]{
Stephen H. Lubow$^{1}$\thanks{E-mail: lubow@stsci.edu}
and
Gordon I. Ogilvie$^{2}$ \thanks{E-mail: gio10@cam.ac.uk}
\\
$^1$Space Telescope Science Institute, 3700 San Martin Drive, Baltimore, MD 21218, USA\\
$^2$Department of Applied Mathematics and Theoretical Physics,
University of Cambridge, Centre for Mathematical Sciences,\\
Wilberforce Road, Cambridge CB3 0WA, UK
}

\pubyear{2024}

\begin{document}
\label{firstpage}
\pagerange{\pageref{firstpage}--\pageref{lastpage}}
\maketitle

\begin{abstract}
The classical Laplace surface defines the location of circular
particle orbits that do not undergo nodal precession around a
planet with some obliquity. Close to the planet the surface coincides with the equator
of the planet, while far from the planet it coincides with the orbital plane of the planet.
We determine the shape of the Laplace
surface of a circumplanetary disc that results from accretion of
circumstellar gas and experiences the effects of gas pressure,
self-gravity, and viscosity, as well as the gravitational effects due
to the planetary spin and the star. We apply the linear theory of
warped discs in the wavelike regime for a small-obliquity planet such
as Jupiter.  As a result of dissipation, a disc that begins slightly
away from its Laplace surface will evolve to it. Because of pressure
effects in typically warm circumplanetary discs, the disc is highly
flattened compared with the classical Laplace surface, meaning that it
is much less warped but still significantly tilted. For the case of
Jupiter, the disc does not align anywhere with the equator of the
planet. For such discs, the effects of self-gravity and viscosity on 
warping  are typically small. The disc tilt is intermediate between
the planet's equatorial and orbital planes.  Circumplanetary discs that are much
cooler than expected can undergo warping and
alignment with the planet's equator at small radii.  The results have
implications for the orbital evolution of satellites in the solar
system that are observed to be somewhat aligned with their classical
Laplace surface.
\end{abstract}

\begin{keywords}
protoplanetary discs --
planets and satellites: formation --
celestial mechanics --
hydrodynamics
\end{keywords}



\section{Introduction}
Gas giant planets are formed from accretion within gaseous circumstellar discs. 
In the core-accretion model, during the late stages of gas-giant planet formation, the planet undergoes runaway
gas accretion \citep{Pollack1996, Lissauer2009, Dangelo2010}. In this phase, the planet is compact and can accept gas at the rate
at which it is supplied by the circumstellar disc. Circumplanetary discs likely form in this phase.
Such planets exert strong tidal forces that are capable of opening a gap in the circumstellar disc
about their orbits \citep{Goldreich1980, Papaloizou1984}. However, these torques do not fully inhibit the flow of gas into the gap region \citep{Artymowicz1996, Gunther2002, Hanawa2010}.
2D simulations suggest that circumplanetary discs can arise from the inflow of circumstellar gas
into the Hill sphere of the planet \citep{Kley1999, Lubow1999}. 
Such material carries angular momentum about  the planet
and can result in the formation of a disc.  In the case of Jupiter, the circumstellar disc thickness is comparable in size
to its Hill sphere. Consequently, the flow into the gap region can involve complex 3D motions that
carry mass and angular momentum onto the circumplanetary disc \citep{Ayliffe2009, Morbidelli2014, Szulagyi2016}. For some  models involving Jupiter,  the inflow results in  an extended rotating planetary envelope, rather than a nearly  Keplerian circumplanetary  disc. The details of the flow
depend on the level of cooling of gas in the gap region 
\citep{Krapp2024}.

The existence of the regular coplanar systems of satellites around
Jupiter and Saturn is suggestive of formation within a circumplanetary disc \citep{Canup2002, Mosqueira2003}.
The Galilean moons are in a nearly coplanar resonant configuration
that can be explained as the outcome of migration within a circumjovian disc \citep{Peale2002}.
PDS 70 is a young T Tauri star that is surrounded by a disc with a partially depleted central cavity (a pre-transitional disc) that contains planets \citep{Hashimoto2012,Haffert2019}.
Recent ALMA observations of system PDS 70 
have provided a  detection
of a circumplanetary disc around planet PDS 70 c
\citep{Benisty2021}. There is also some observational evidence from the VLT for a circumplanetary disc around PDS 70 b \citep{Christiaens2019}.

Consider satellites that are on circular orbits about a planet.
The satellites are subject to gravitational torques exerted by the central planet and by the star about which the planet orbits. 
For a spherical planet, there are no torques on the satellite exerted by the planet.
In that case, the orbital plane of the planet about the star defines a flat surface on which there is no torque and thus no  precession of the satellite.
If the planet is non-spherical as a result of its spin, then 
the planet can exert a precessional torque on a satellite whose orbit is misaligned with the planet's equator. If the satellite lies on a non-precessing
orbit, then the sum of the torques exerted by planet and star vanishes. If the planet has a non-zero obliquity, i.e., if its equator 
is tilted relative to its orbit about the star, then the vanishing-torque condition defines a set
of satellite orbits that lie on a curved surface called the Laplace surface. Close to the planet, this classical Laplace surface
coincides with the equatorial plane of the planet. Far from the planet, but well within its Hill radius, the surface
coincides with the orbital plane of the planet. Near an intermediate radius, called the Laplace radius,
there is a smooth change in tilt between these two orbital orientations.
The Galilean moons are found to lie
on orbits that are roughly coplanar with the equator of Jupiter that has an obliquity of $3.13^{\circ}$. Their orbits are within about $0.5^\circ$ of the classical Laplace surface and lie interior to the Laplace radius of Jupiter. In the presence of other forces such as gas pressure, the Laplace surface of a gaseous disc, for which the disc
is non-precessing, may deviate from the classical Laplace surface.

A key issue is how a satellite achieves its orbit near the classical Laplace surface.
It could have formed there if the  midplane of its natal gaseous circumplanetary disc coincided
with the classical Laplace surface \citep[e.g.,][] {Tremaine2009, Tremaine2023}. 
The circumplanetary disc is expected follow the orbits of particles, if the disc
is sufficiently cool. The disc is also dissipative and would be expected to evolve towards
the Laplace surface, where the gravitational torques cancel out. However, there is also a hydrodynamic torque associated with gas pressure and viscosity in the disc, which increases with gas temperature or equivalently disc thickness. Owing to the effects of pressure, that torque 
acts to spread and reduce a warp in the disc. 
Self-gravity can also modify the Laplace surface. \cite{Ward1981} calculated the effects of disc self-gravity in altering
the Laplace surface in an effort to explain the orbital tilt of Iapetus that is far ($\sim 8^{\circ}$) from the classical Laplace surface of Saturn. He also recognized the potential importance of gas pressure, but did not analyse its effects.
\cite{Zanazzi2018a} studied the warp in a precessing circumstellar disc involving a misaligned stellar spin axis within a binary star system and included the effects of disc gas pressure and viscosity.
In this paper we determine the Laplace surface of a circumplanetary disc by requiring it to be non-precessing and accounting for the effects of pressure, self-gravity, and viscosity. 

The outline of the paper is as follows.
Section \ref{sec:eqs} derives the differential equations for the Laplace surface of a circumplanetary disc based on the linear theory of warped discs applied to a small-obliquity planet. We consider applications to a
circumjovian disc.
Section \ref{sec:cold} shows that
 the Laplace surface of a disc with typical parameters is dominated by the effects of pressure and cannot be described by a model based on particle orbits.  Section \ref{sec:warm} describes the Laplace surface of a typically warm, non-self-gravitating, inviscid circumplanetary disc. Section \ref{sec:cool} describes  results for cooler, non-self-gravitating, inviscid discs.  Section \ref{sec:sg} analyses the effects of disc self-gravity on inviscid warm discs. Section \ref{sec:visc} describes  the effects of viscosity on warm discs.
 Section \ref{sec:summary} contains a discussion and summary. 

\section{Disc Equations and Model Parameter Values}
\label{sec:eqs}
We consider a system consisting of a star, a planet, and a circumplanetary disc.
  We assume that the planet and disc
are on circular orbits. 
We determine the disc tilt vector (disc angular momentum unit vector) 
$\bm{\ell}$ as a function of the distance $r$ from the planet centre. 
 
 We use the linearized warped disc evolution equations that apply in the wavelike regime
 in which $\alpha < H/r$, where $\alpha$ is the standard viscosity parameter and $H$ is the disc thickness.
 They are also valid in the diffusive regime, provided that $\alpha\ll1$.
 The time-dependent equations for the angular momentum and torque of the circumplanetary disc with surface density $\Sigma(r)$ 
are given by
\begin{align}
&\Sigma r^2 \Omega \frac{\partial \bm{\ell}}{\partial t} =\frac{1}{r}\frac{\partial \bm{F}}{\partial r}+ \Sigma  \bm{T}  \label{dldt}\\
& \text{and} \nonumber \\ 
& \frac{\partial \bm{F}}{\partial t} - \omega_{\rm a} \bm{\ell} \times \bm{F}+\alpha \Omega  \bm{F} =\frac{1}{4} \Sigma H^2 r^3 \Omega^3 \frac{\partial \bm {\ell}}{\partial r} \label{dFdt}
\end{align}
\citep{Papaloizou1995a, Papaloizou1995, Lubow2000}. Here $\bm{F}(r,t)$ is the angular momentum flux (per radian in azimuth) that radially communicates the effects of the disc warp through pressure and viscosity, $\bm{T}(r,t)$ is the gravitational torque per unit mass on the disc caused by the planet-star system, and $\omega_{\rm a}(r)$ is the apsidal precession rate of slightly eccentric orbits.
The angular velocity $\Omega(r)$ of the disc is otherwise taken to be Keplerian, $\Omega=(GM_{\rm p}/r^3)^{1/2}$.
  We typically take as boundary conditions that the angular momentum flux vanishes
  at the disc edges. That is
\begin{equation}
\bm{F}(r_{\rm in}, t) = \bm{F}(r_{\rm out}, t) = \mathbf{0} \label{BCF}
\end{equation}
at the inner and outer disc radii $r_{\rm in}$ and $r_{\rm out}$, respectively.
 
  We take
 $\bm{n}_{\rm p}$ and $\bm{n}_{\rm s}$ to be  unit vectors parallel to the spin of the planet and the orbital angular momentum of the planet about the star, respectively, which are assumed to be fixed in time.
Equations 5.66 and 5.72 of \cite{Tremaine2023}  provide the gravitational torque per unit mass on the disc due  to the planet  and the star:
\begin{equation}
\bm{T}  = r^2 \Omega  [\omega_{\rm p} (r) (\bm{\ell} \cdot \bm{n}_{\rm p}) (\bm{\ell} \times \bm{n}_{\rm p}) + \omega_{\rm s} (r) (\bm{\ell}\cdot \bm{n}_{\rm s}) (\bm{\ell} \times \bm{n}_{\rm s}) ] ,\label{T0}
\end{equation}
where 
\begin{equation}
\omega_{\rm p} (r)  =  \frac{3 (G M_{\rm p})^{1/2} J_2 R_{\rm p}^2}{2 r^{7/2}}
\end{equation}
and
\begin{equation}
\omega_{\rm s} (r)  =  \frac{3  G^{1/2}  M_{\rm s} r^{3/2}}{ 4 M_{\rm p}^{1/2} a_{\rm s}^3},
\end{equation}
in which $-\omega_{\rm p}$ and $-\omega_{\rm s}$ are the nodal precession rates of slightly inclined orbits due to the planet and star, respectively.
The planet has mass  $M_{\rm p}$, equatorial radius $R_{\rm p}$, and dimensionless quadrupole moment $J_2$. The star has mass  $M_{\rm s}$ and is orbited by the planet at radius 
$a_{\rm s}$.  

We assume the level of misalignment between $\bm{n}_{\rm p}$ and $\bm{n}_{\rm s}$
is small so that $\bm{\ell} \cdot \bm{n}_{\rm p}$ and $\bm{\ell} \cdot \bm{n}_{\rm s}$ can be approximated as unity in Equation (\ref{T0}).
We take the disc tilt to be stationary, as would be required for it to lie on the Laplace surface, so that $\partial \bm{l}/\partial t = \bm{0}$ and
$\partial \bm{F}/\partial t = \bm{0}$.  We first consider an inviscid disc, $\alpha=0$, and ignore the effects of disc self-gravity.
As in the derivation of the Laplace surface for particles,  we consider the case that
$\bm{n}_{\rm p}$, $\bm{n}_{\rm s}$, and $\bm{\ell}$ lie in a common plane. Without loss of
generality, we apply a Cartesian coordinate system and take 
the $x-z$ plane to contain vectors
$\bm{n}_{\rm p}$ and $\bm{n}_{\rm s}$, with the $z$-direction parallel to the total angular momentum 
of the planet. To first order 
in the tilt angle $0< \beta \ll 1$ of vector $\bm{\ell}$ away from the $z$-direction, and similarly for $\bm{n}_{\rm p}$ and $\bm{n}_{\rm s}$, we 
have
\begin{align}
 \bm{\ell} = (\beta, 0, 1), \qquad
 \bm{n}_{\rm p}  = (\beta_{\rm p}, 0, 1), \qquad \text{and} \qquad
 \bm{n}_{\rm s}  = (\beta_{\rm s}, 0, 1). \label{betaapprox}
\end{align}
The gravitational torque is in the $y$-direction and is then approximately given by
\begin{equation}
T_y  = -r^2 \Omega [ \omega_{\rm p} (\beta - \beta_{\rm p})  + \omega_{\rm s} (\beta - \beta_{\rm s}) ]. \label{Ty}
\end{equation}
 We apply these assumptions to Equations (\ref{dldt}) and (\ref{dFdt}) to obtain
\begin{equation}
  \Sigma r^2 \Omega [ \omega_{\rm p} (\beta - \beta_{\rm p})  + \omega_{\rm s} (\beta - \beta_{\rm s}) ] = \frac{1}{r}\frac{\dd F_y}{\dd r}  \label{Ty1} \\
\end{equation}
and
\begin{equation}
  \omega_{\rm a} F_y =\frac{1}{4} \Sigma H^2 r^3 \Omega^3 \frac{\dd \beta}{\dd r},  \label{Fy1}
\end{equation}
where the apsidal precession rate for orbits of small tilt is $ \omega_{\rm a}(r) =  \omega_{\rm p}(r) + \omega_{\rm s}(r)$.

Throughout this paper, we consider a disc model with constant disc aspect ratio $h=H/r$ and  $\Sigma(r) = \Sigma_0/r$ with constant $\Sigma_0$.
 Gas pressure can also contribute to $\omega_{\rm a}$. However, for the disc model we consider, the apsidal precession rate due to pressure vanishes. Near disc edges there are deviations from this density distribution due to disc tapering, resulting in some apsidal
contributions that could affect the tilt. But the tilt changes are
mainly confined to the disc tapering region
and can be shown to be small.
 For alternative disc models which we do not consider in this paper, pressure could contribute to $\omega_{\rm a}$, but these effects on the disc tilt
are typically small and do not change our main conclusions.

For the disc model we consider, Equations (\ref{Ty1}) and (\ref{Fy1}) are respectively written in dimensionless form as
\begin{equation}
\frac{\beta - \beta_{\rm p}}{x^3} + (\beta-\beta_{\rm s}) x^2 = p \frac{\dd f}{\dd x} \label{beta}
\end{equation}
and
\begin{equation}
\left(\frac{1}{x^3} + x^2 \right) f = p \frac{\dd \beta}{\dd x} \label{f},
\end{equation}
where
\begin{align}
& x = r/R, \\ \label{xdef}
& R = \left( 2 J_2 R_{\rm p}^2 a_{\rm s}^3 \left(\frac{M_{\rm p}}{M_{\rm s}} \right) \right)^{1/5},\\ 
& p= \frac{2^{2/5}}{3} h  \left(\frac{a_{\rm s}^2}{R_{\rm p}^2 J_2} \right)^{3/5} \left(\frac{M_{\rm p}}{M_{\rm s}} \right)^{2/5},\label{pdef} \\ 
 & f = \frac{2 F_y}{\Sigma_0 h G M_{\rm p}}. 
\end{align}
Notice that our normalization radius $R=2^{1/5} r_{\rm L}$, where  $r_{\rm L}$
is the Laplace radius given by Equation 5.74 of \cite{Tremaine2023}.
The boundary conditions given by Equation (\ref{BCF})
become
\begin{equation}
f(x_{\rm in}) = f(x_{\rm out}) =0. \label{BCf}
\end{equation}

The dimensionless parameter $p$ measures the importance of gas pressure in determining the shape of the Laplace surface. The bending-wave propagation speed is $c_\text{s}/2$ 
where $c_{\rm s} = H \Omega$ is the isothermal gas sound speed \citep{Papaloizou1995a}.
For the case of a circumstellar disc that is tidally perturbed by a binary companion, 
the condition for a nearly flat disc (weak disc warping) due to the effects of pressure is that the sound crossing time (similar to the bending-wave crossing time) is shorter than 
the nodal precession period \citep{Papaloizou1995,Larwood1996}.
In the current case, there are
two nodal precession frequencies: $-\omega_{\rm p}$
and $-\omega_{\rm s}$. Parameter $p$ can be expressed as
\begin{equation}
\f{1}{p} = \frac{2r}{c_{\rm s}} \,  \omega_{\rm p}^{3/5} \omega_{\rm s}^{2/5} =\frac{2}{h}\frac{\omega_{\rm p}^{3/5}\omega_{\rm s}^{2/5}}{\Omega}. 
\end{equation}
The bending-wave crossing time is $2 r/c_{\rm s}$ and the other factors involve a combination of the two precession frequencies.  The RHS  of this equation  therefore compares the crossing time with a weighted geometric mean of the two precession rates. Note that $p$ is 
independent of radius for the constant disc aspect ratio adopted; more generally, $p\propto h$. Condition $p > 1$ generalizes the criterion  for weak warping in the single torque case to the current case
of two precessional torques.

In this paper, we consider circumjovian  discs having various model parameters. They are given in Table~\ref{table}.
\begin{table*}
\begin{center}
\begin{tabular}{l c c c c c c c c c  c c l}
\hline
 Model & $h$ & $r_{\rm out}/R_{\rm H}$  & \text{self-gravity}   & \text{viscosity}   \\
 \hline
\hline
A  &  0  & 0.4 & No & No   \\  
B &  0.1  & 0.4 & No & No \\
C &   0.3  &  1 &  No & No \\
D &  0.001  & 0.4 & No& No  \\
E &  0.1  & 0.4 & Yes & No   \\ 
F &  0.1  & 0.4 & No & Yes   \\ 
\hline
\end{tabular}
\end{center}
 \caption{Parameters for  circumjovian disc models. $h$ is the disc aspect ratio (assumed constant), $r_{\rm out}/R_{\rm H}$ 
 is the ratio of the disc outer radius to the Hill radius of Jupiter. For all of these models, the disc inner
 radius is the radius of Jupiter, the dimensionless quadrupole moment  is $J_2=0.015$, and the obliquity of Jupiter is $3.13^\circ$.
All models have zero angular momentum flux as a boundary condition at the disc inner and outer radii, except in the case of Model C that has
the outer disc edge tilt aligned with the orbital plane of the circumstellar disc.}
\label{table}
\end{table*}

\section{Laplace surface of a cold disc}
\label{sec:cold}

In the case of a cold disc ($h=0$), Model A, we have that $p=0$. We apply Equations (\ref{beta}) and (\ref{f})
to find that the angular momentum flux $f=0$ and that the disc tilt is given by
\begin{equation}
\beta_{\rm cold}(x) = \frac{\beta_{\rm p}+ x^5 \beta_{\rm s}}{1+x^5}. \label{betacold}
\end{equation}
Tilt $\beta_{\rm cold}$ is equal to the tilt given
by Equation 5.76 of \cite{Tremaine2023} for a particle disc in the small-tilt approximation.
(The tilt relative to the planet is obtained by setting $\beta_{\rm p}=0$.)
For the cold disc solution to be applicable in the presence of some pressure, 
we require $p \ll 1$
or 
\begin{equation}
h \ll \frac{3}{2^{2/5}}  \left(\frac{R_{\rm p}^2 J_2}{a_{\rm s}^2} \right)^{3/5} \left(\frac{M_{\rm s}}{M_{\rm p}} \right)^{2/5}. \label{hcold}
\end{equation}
We consider the case of Jupiter and adopt the current values of parameters, including 
$J_2=0.015$. Equation (\ref{hcold}) implies
 that $h \ll 4 \times 10^{-5}$.
Such small values of $h$ are unrealistic for a circumplanetary disc for which we expect $h \gtrsim 0.1$ \citep[e.g.,][]{Martin2023}.
In that case, we have that $p \gtrsim 2.5 \times 10^3$.   This suggests
that the disc does not lie on the classical particle Laplace surface because of the dominance
of pressure effects.

\section{Laplace surface of a warm gaseous disc}
\label{sec:warm}

The large value of $p$ for Jupiter indicates that the disc is in good radial communication because pressure effects  play the dominant role in determining the shape of the circumjovian Laplace surface.
Instead of regarding $p$ as a small parameter as in Section \ref{sec:cold}, we instead assume $p$ to  be large.
We define $\epsilon = 1/p \ll1$ and expand the solution of Equations (\ref{beta}) and (\ref{f}) as
\begin{equation}
\beta =  \beta_0 + \epsilon \beta_1 + \cdots \label{betaexpansion}
 \end{equation}
 and \begin{equation}
 f =  f_0 + \epsilon f_1 + \cdots. \label{fexpansion}
 \end{equation}
 We regard $x$ and $\beta$ as order-unity quantities in the expansion in $\epsilon$.
 We have already taken  $\beta$ to be small in Equation (\ref{betaapprox}). We assume
 now that $\epsilon \ll \beta \ll 1$.
 
 To lowest order,
 \begin{align}
  \frac{\dd \beta_0 }{\dd x} = 0  \qquad\text{and} \qquad
  \frac{\dd f_0 }{\dd x}  =0. \label{dbeta0dx} 
 \end{align}
 From Equation (\ref{BCf}), we have that 
 \begin{equation}
 f_0(x)=0.
 \end{equation}
 
 In next order, we integrate Equation (\ref{beta}) from $x_{\rm in}$ to $x_{\rm out}$ and
 apply  Equation (\ref{BCf}) to obtain
 \begin{equation}
 \beta_0 = \frac{\left( -  \frac{\beta_{\rm p}}{2 x^2} +   \frac{\beta_{\rm s} x^3}{3} \right)  \bigg\rvert_{x_{\rm in}}^{x_{\rm out}}}
 { \left(  -  \frac{1}{2 x^2}+ \frac{x^3}{3} \right)  \bigg\rvert_{x_{\rm in}}^{x_{\rm out}}}. \label{beta0}
 \end{equation}
 We note that for $x_{\rm in} \ll 1 \ll x_{\rm out}$, Equation (\ref{beta0}) reduces to
 \begin{equation}
  \beta_0 \simeq  \frac{ 3 \beta_{\rm p} +   2   x_{\rm in}^2 x_{\rm out}^3 \beta_{\rm s}}
 {   3 +  2 x_{\rm in}^2 x_{\rm out}^3}.  \label{beta0xixo}
  \end{equation}
We then have 
 \begin{equation}
 f_1(x) =  \left( -\frac{(\beta_0-\beta_{\rm p}) }{2 x^2} +\frac{ (\beta_0-\beta_{\rm s}) x^3}{3} \right)  \bigg\rvert_{x_{\rm in}}^{x} \label{f1}
 \end{equation}
and 
\begin{equation}
\beta_1(x) = 0.
\end{equation}
In next order in Equation (\ref{f}),
 \begin{equation}
 \beta_2(x) =  \int \left(\frac{1}{x^3}+x^2 \right) f_1(x) \, \dd x + c,
 \end{equation}
 where $c$ is a constant of integration.
 From Equation (\ref{beta})  together with Equation (\ref{BCf}),
 we have an integral constraint
 \begin{equation}
 \int_{x_{\rm in}}^{x_{\rm out}} \left ( \frac{1}{x^3} +  x^2 \right)   \beta_2(x) \, \dd x =0 \label{cint}
 \end{equation}
that determines $c$.
We then calculate
\begin{equation}
    \beta(x) \simeq \beta_0 + \epsilon^2 \beta_2(x). \label{betaexp} 
\end{equation}
Constant tilt $\beta_0$ is given by Equation (\ref{beta0}). The $\beta_2$ term
has a contribution due to the disc warp.
Equation (\ref{betastdexp}) provides an analytic expression for $\beta_2(x)$.

These expansions hold provided that $x$ is an order-unity quantity throughout the disc or equivalently that $x_{\rm in}$ and $x_{\rm out}$ are order-unity quantities. For $x$ small, Equations (\ref{beta}) and (\ref{f}) imply that the above expansions break down for
 $x^3 \partial /\partial x \sim \epsilon$. 
At the inner boundary this occurs for $x_{\rm in} \sim \epsilon^{1/2} = 1/p^{1/2}$.
 For $x$ large, the expansions break down for
 $x^{-2} \partial /\partial x \sim \epsilon$.
At the outer boundary this occurs for $x_{\rm out} \sim 1/\epsilon^{1/3} = p^{1/3}$.
We expect that the above solution of nearly constant $\beta$ holds provided 
that $ x_{\rm in} \gtrsim 1/p^{1/2}$ and  $x_{\rm out} \lesssim p^{1/3}$.  These limits are described as inner and outer
disc Laplace radii
in Section \ref{sec:cool}, which also describes the behaviour outside these limits.

We consider the case of Jupiter and take the inner disc radius  to be the current equatorial radius $R_{\rm J}$ of Jupiter
and obtain
that $x_{\rm in}  \approx 0.03 >  1/p^{1/2} \approx 0.02$.  
 We take the outer disc radius  to be $0.4 R_{\rm H}$, where $R_\text{H}=(M_{\rm p}/3M_{\rm s})^{1/3}a_{\rm s}$ is the Hill radius; this estimate applies to a thin circumplanetary disc that is truncated by orbit crossings \citep{Martin2011} due an evection resonance \citep{Kisare2024}. We then have that $x_{\rm out} \approx 9.2 <  1/\epsilon^{1/3} \approx 13.5$. So the
nearly constant tilt solution should be at least marginally satisfied throughout the disc. 
For the case of a circumjovian disc in Model B with $h=0.1$ and Jupiter's obliquity of $3.13^{\circ}$, we evaluate $\beta_0$
from Equation (\ref{beta0}) and $\beta_2$ 
 from Equation (\ref{betastdexp}) to obtain from Equation (\ref{betaexp})
\begin{equation}
\beta(x) \simeq 1.016 + 2.050 \times 10^{-8} x^{-4} - 4.571 \times 10^{-5} x^{-2}
-1.973 \times 10^{-7} x + 3.047 \times 10^{-5} x^3 - 1.922 \times 10^{-8} x^6, \label{betastd}
\end{equation}
where $\beta$ is in degrees. 

\begin{figure}
	\includegraphics[width=0.8\textwidth]{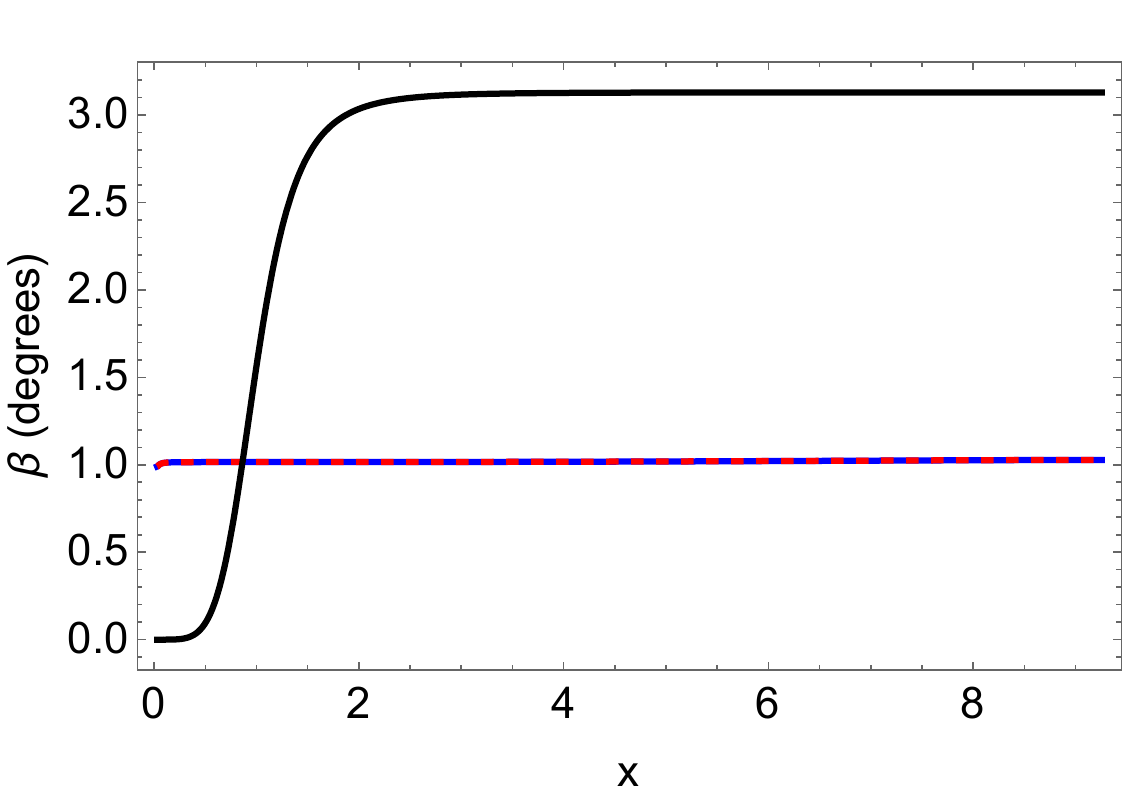}\\
    \includegraphics[width=0.8\textwidth]{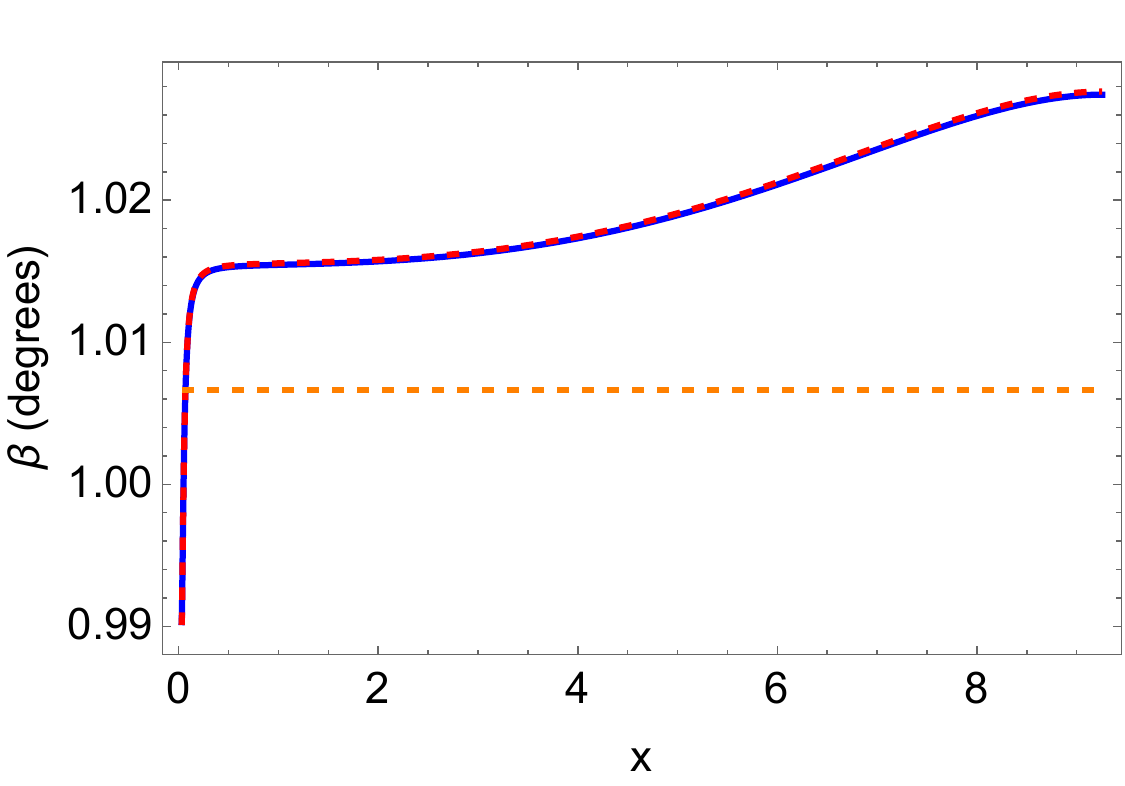}
    \caption{Plot of $\beta(x)$, the local tilt angle of the Laplace surface relative to the equatorial plane of Jupiter, as a function of dimensionless distance from the centre of Jupiter.
    The black line is for a cold circumplanetary disc and corresponds to the classical particle Laplace surface, Model A. All other lines are for the warm disc Model B. The blue line is based on a numerical solution
    to Equations (\ref{beta})--(\ref{BCf}). 
    The overlapping dashed red line is obtained from the  analytic expression in Equation (\ref{betastd}).   The lower plot is a closeup of the upper plot for comparing the  numerical and analytic results. In addition it contains a plot of the lowest order analytic approximation given by $\beta_0$ in Equation (\ref{beta0}) shown as the dashed orange line. $x=1$ corresponds to about $33 R_{\rm J}$.}    
    \label{fig:betastd}
\end{figure}

Figure~\ref{fig:betastd} plots $\beta(x)$ for the Laplace surface around Jupiter, where $x \simeq r/(33 R_{\rm J})$.
The black line plots  the case of a cold circumplanetary disc (Model A) obtained by using Equation (\ref{betacold}).  
The solid blue line plots 
the nearly flat (but tilted) numerical solution to Equations (\ref{beta})--(\ref{BCf}) for a warm disc with properties given by Model B, $h=0.1, J_2=0.015, r_{\rm in} =R_{\rm J}$, and $r_{\rm out}=0.4 R_{\rm H}$.  The lowest order approximation for the tilt, $\beta_0$ given by Equation (\ref{beta0}), is plotted as the orange dashed line in the bottom panel.   The more accurate analytic approximation   given by Equation (\ref{betastd}) with $h=0.1$ is plotted as the red dashed line. 
It agrees well with the numerical solution. The disc is not aligned with either the planet's equator or the circumstellar disc but is closer to the former.

\begin{figure}
   \includegraphics[width=0.8\textwidth]{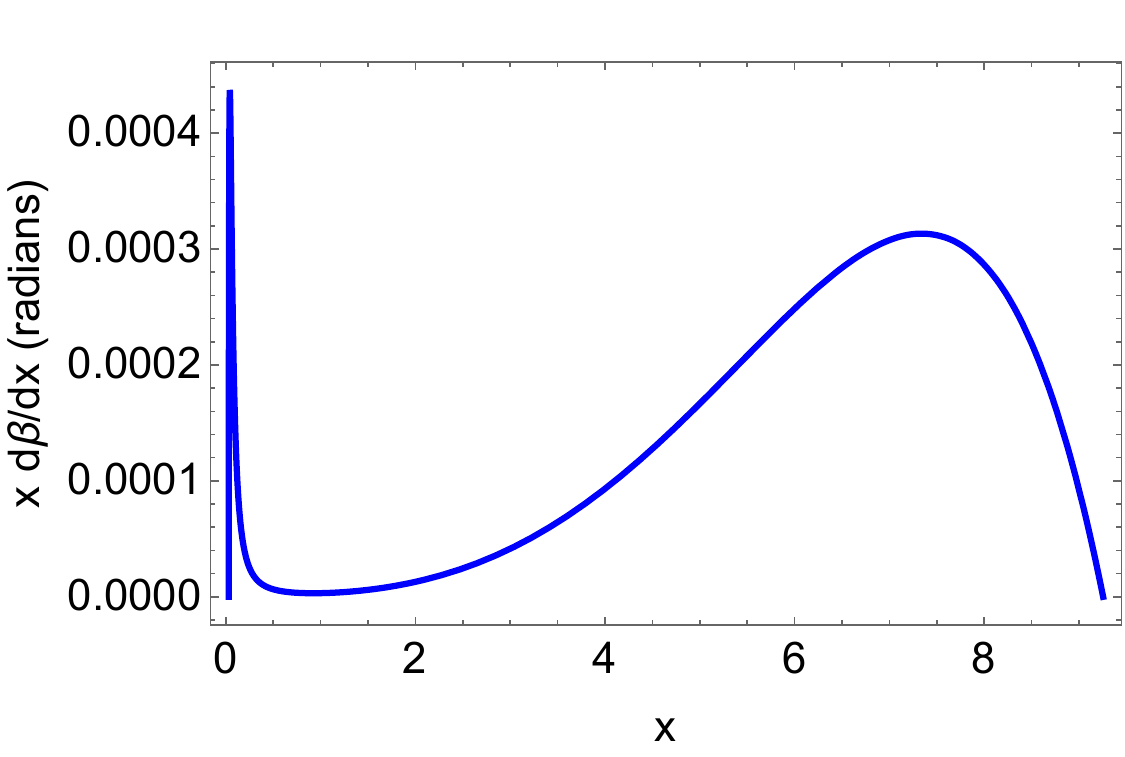}
    \caption{Plot of the warp, $x\, \dd \beta/\dd x$, as a function of dimensionless distance from the centre of Jupiter for Model B plotted as the the blue line in Figure~\ref{fig:betastd}. }   
    \label{fig:warpstd}
\end{figure}

Figure~\ref{fig:warpstd} plots $x \, \dd \beta/\dd x$ in radians, a measure of the disc warp, as a function of $x$.
As expected, this function is zero at the inner and outer boundaries as required by the boundary conditions given in Equation (\ref{BCf}). It is numerically small everywhere  and much less than $\sqrt{h} \approx 0.3$,
suggesting that nonlinearities do not play a role \citep{Ogilvie2006}.

In the  case of Jupiter, the disc inner edge may actually be somewhat larger than Jupiter's current radius $R_{\rm J}$ as we have assumed in Model B. The proto-Jupiter radius may have been somewhat larger $\sim 1.5 R_{\rm J} -1.8 R_{\rm J}$ \citep{Lissauer2009}.
Also, \cite{Batygin2018} estimates that  $r_{\rm in }  \sim 5 R_{\rm J}$,
if the inner disc is truncated by the planet's magnetic field. We  consider how $\beta$
varies as a function of disc inner radius 
with all other parameters taken to be the same.
For such larger values of  $r_{\rm in }$ the disc should be even flatter.

\begin{figure}
   \includegraphics[width=0.8\textwidth]{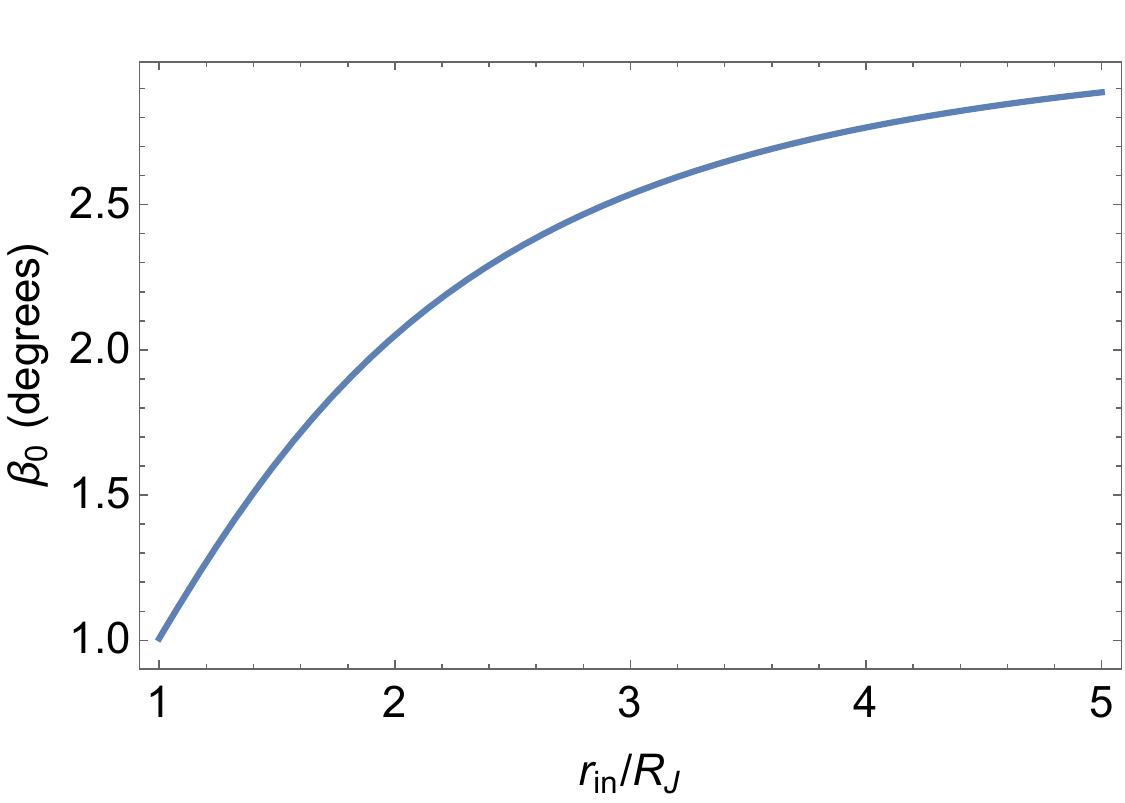}
    \caption{Plot of the tilt angle $\beta_0$ given by Equation (\ref{beta0}) as a function of disc inner radius normalized by Jupiter's radius with all other parameters given by Model B. }    
    \label{fig:beta0rin}
\end{figure}

Figure~\ref{fig:beta0rin} plots $\beta_0$ as a function of radius normalized by Jupiter's radius, $r_{\rm in}/R_{\rm J}$, with other parameters that follow Model B.  With increasing disc inner radius, the warp near the disc inner edge weakens and $\beta$ becomes more accurately approximated as $\beta_0$.
Therefore, over this range of disc inner radii, the disc
tilt is nearly constant in radius and nearly equal to $\beta_0$. As seen in Figure~\ref{fig:beta0rin}, the disc tilt varies with inner disc radius from about 1/3 of Jupiter's obliquity to nearly the value of Jupiter's obliquity 
for which the disc is coplanar with the orbit of Jupiter.

A thick circumplanetary disc with $h \gtrsim 0.3$  may extend to the Hill radius and join the circumstellar disc
\citep{Martin2023}.  In addition the disc rotation is somewhat non-Keplerian, an effect we ignore here. 
For this case, we consider a different outer boundary condition
that is $\beta(x_{\rm out}) = \beta_{\rm s} = 3.13^{\circ}$ in Model C.
With this outer boundary condition and the inner boundary condition that $f(x_{\rm in}) =0$, we have from Equations (\ref{beta}) and (\ref{f})
with again $\epsilon=1/p$,
\begin{align}
     \beta_0 = \beta_{\rm s} \label{beta0thick} \qquad\text{and}
    \qquad f_0 = 0.
\end{align}
In next order in $\epsilon$, we have from Equation (\ref{beta}) that
\begin{equation}
f_1(x) = \frac{1}{2}(\beta_{\rm s} - \beta_{\rm p}) 
\left(
\frac{1}{x_{\rm in}^2} - \frac{1}{x^2} 
\right)
\end{equation}
and 
\begin{equation}
\beta_1(x) = 0.
\end{equation}
To next order in Equation (\ref{f}) we obtain
\begin{equation}
\beta_2(x) = \int_{x_{\rm out}}^x 
\left(
\frac{1}{x'^3} + x'^2 
\right) f_1(x') \, \dd x'.
\end{equation}
Again, $\beta_2(x)$ is a polynomial function of $x$ that contains some inverse power terms. 
We then calculate $\beta(x) \simeq \beta_0 + \epsilon^2 \beta_2(x)$.

In this case, the circumplanetary disc tilt is close to the circumstellar disc tilt and is fairly constant in radius. It is in good agreement with the analytic prediction. 
For this disc, the warp is sufficiently small, $x \, \dd\beta/\dd x < 0.006$, that nonlinearities are likely unimportant \citep{Ogilvie2006}.

 \begin{figure}
   \includegraphics[width=0.8\textwidth]{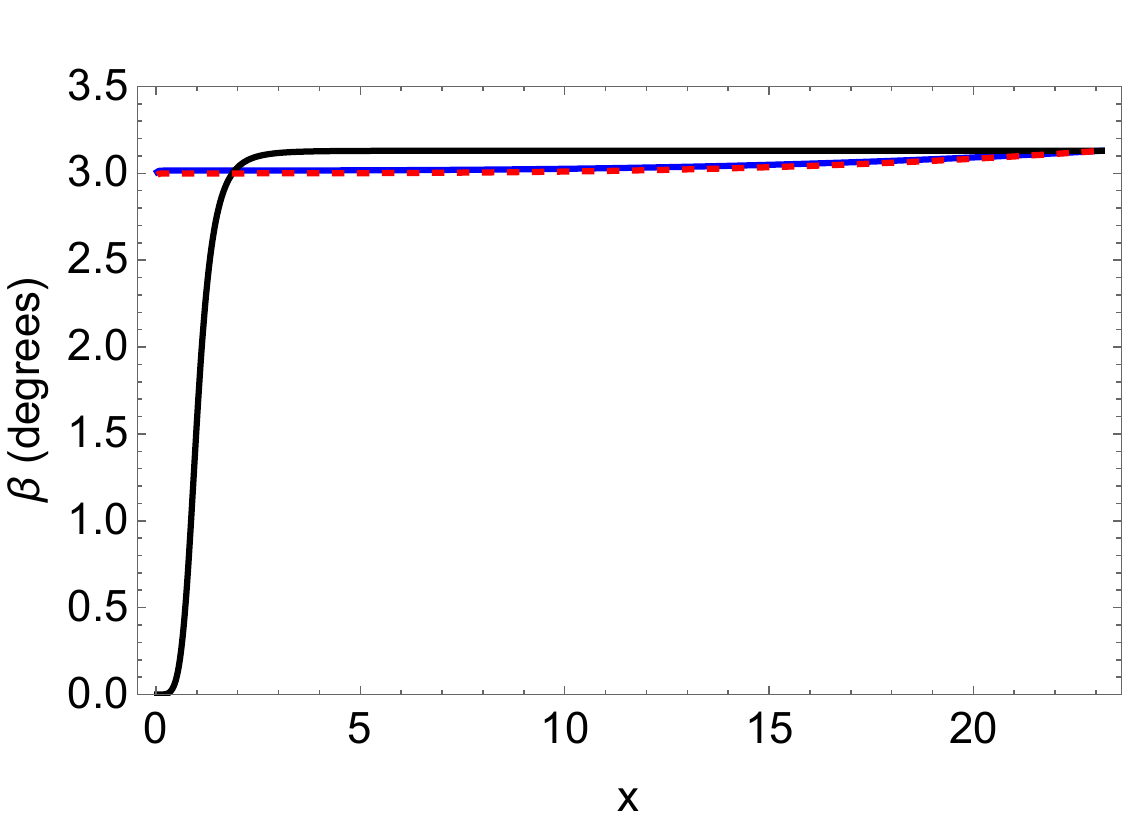}
    \caption{ Same as the upper panel of Figure~\ref{fig:betastd} but for a circumplanetary disc with aspect ratio $h=0.3$
    that extends to the Hill sphere where it is coplanar with the circumstellar disc, Model C. The dashed red line is the analytic
    prediction for a thick disc.}
    \label{fig:betaRH}
\end{figure}


For the expansions to significantly break down near the inner boundary, we require 
\begin{equation}
x_{\rm in} \sqrt{p} = 1.5 \left( \frac{h}{0.1} \, \, \frac{0.015}{J_2} \right)^{1/2} \ll 1. \label{xilim}
\end{equation} 
Therefore, unless the disc is very cool, the inner expansion is at least marginally valid near the inner boundary.
At the outer boundary taken to be at a distance $0.4 R_{\rm H}$, significant breakdown
requires
\begin{equation}
\frac{x_{\rm out}}{p^{1/3}} = 0.7 \left( \frac{0.1}{h} \right)^{1/3} \gg 1, \label{xolim}
\end{equation}
which again is not satisfied unless the disc is very cool.
Notice that Equations (\ref{xilim}) and (\ref{xolim}) are independent of all other parameters, such as $R_{\rm p}$, $M_{\rm p}$, $a_{\rm s}$, etc.
The flat disc tilt profile is insensitive to changes in disc and planet
parameters. The tilt value does depend on the inner and outer boundary conditions as seen in Figure~\ref{fig:beta0rin} and
in comparing Figures~\ref{fig:betastd} and \ref{fig:betaRH}.

The conclusion is that a circumjovian disc is nearly flat throughout
under a wide variety of conditions, unless it is very cool, $h \ll 0.1$.
In no warm disc case we consider does the disc align with the equator of the planet. The level of misalignment is a significant fraction of the planet's obliquity.

\section{Laplace surface of a cool gaseous disc}
\label{sec:cool}

In the previous section, we found that a warm circumjovian disc does not contain
the analogue of the Laplace radius where its tilt undergoes a transition from
an orientation along the equator of the planet to an orientation along the planet's orbital plane. In this section, we consider a `cool' disc in which tilt transitions do occur.
Unlike the cold disc considered in Section \ref{sec:cold}, however, we assume that
 the cool disc still has $p\gg1$. In this case, there is a broad flat region of the disc at intermediate radii
 because the disc is in overall good radial communication by pressure.
 However, warping can occur in the inner and/or outer regions of the disc where tidal effects are strong.
 There are then up to two analogues of the Laplace radius at which tilt transitions occur.
 As a result of the inner warp, the inner region of the disc can align with the equator of the planet, if the disc extends
 to smaller radii than the inner Laplace radius.
As a result of the outer warp, the outer region of the disc can align with the orbital plane of the planet,  if the disc extends
 beyond the outer Laplace radius.

A way to derive the approximate solution of Equations (\ref{beta}) and (\ref{f}) for small $r$ is to neglect the $x^2$ terms contributed by the outer quadrupole due to the star. Then there are two solutions:
\begin{align}
  f=\beta-\beta_\text{p}=\exp\left(-\f{1}{2px^2}\right)\qquad\text{and}\qquad
  -f=\beta-\beta_\text{p}=\exp\left(+\f{1}{2px^2}\right),
\end{align}
each of which can be scaled by an arbitrary factor, and each of which tends to a constant tilt at large $r$. We apply the boundary condition  given by Equation (\ref{BCf}) so that the torque (and therefore $f$) vanishes at $x=x_\text{in}$. The relevant solutions are
\begin{equation}
  f=C_\text{in}\left[\exp\left(-\f{1}{2px^2}\right)-\gamma_\text{in}\exp\left(+\f{1}{2px^2}\right)\right],\qquad
  \beta-\beta_\text{p}=C_\text{in}\left[\exp\left(-\f{1}{2px^2}\right)+\gamma_\text{in}\exp\left(+\f{1}{2px^2}\right)\right],  \label{betain}
\end{equation}
where
\begin{equation}
  \gamma_\text{in}=\exp\left(-\f{1}{px_\text{in}^2}\right) \label{gammain}
\end{equation}
is less than (and possibly very much less than) unity.

Similarly, an approximate solution for large $r$ can be derived by neglecting the $x^{-3}$ terms contributed by the inner quadrupole due to the planet:
\begin{equation}
  f=C_\text{out}\left[\exp\left(-\f{x^3}{3p}\right)-\gamma_\text{out}\exp\left(+\f{x^3}{3p}\right)\right],\qquad
  \beta_\text{s}-\beta=C_\text{out}\left[\exp\left(-\f{x^3}{3p}\right)+\gamma_\text{out}\exp\left(+\f{x^3}{3p}\right)\right], \label{betaout}
\end{equation}
where
\begin{equation}
  \gamma_\text{out}=\exp\left(-\f{2x_\text{out}^3}{3p}\right)
\end{equation}
is less than unity.
For the case of the warm circumjovian disc (Model B) plotted in Figure~\ref{fig:betastd}, we have that $\gamma_\text{in}\approx0.64$ and $\gamma_\text{out}\approx0.81$. For a cooler disc, $\gamma_\text{in}$ and/or $\gamma_\text{out}$ could be much less than unity.


Note that, in dimensional terms, these elementary solutions have the general form
\begin{equation}
  \pm f=\beta_\text{rel}\propto\exp\left(\pm\int\f{2\omega_\text{a}}{c_\text{s}}\,\dd r\right),
\end{equation}
where $\beta_\text{rel}$ is the tilt of the disc relative to the source of the precession, $\omega_\text{a}(r)$ is the apsidal precession rate and $c_\text{s}(r)$ is the sound speed. For $h=H/r=\text{constant}$, the exponential function can also be written as
\begin{equation}
  \exp\left(\pm\f{2}{h}\int\f{\omega_\text{a}}{\Omega}\,\dd\ln r\right).
\end{equation}
For the inner quadrupole ($\omega_\text{a}/\Omega\propto r^{-2}$) this gives $\exp[\mp(1/h)(\omega_\text{a}/\Omega)]$. For the outer quadrupole ($\omega_\text{a}/\Omega\propto r^3$) it gives $\exp[\pm(2/3h)(\omega_\text{a}/\Omega)]$. 

The inner approximate solution\footnote{These inner and outer approximate solutions are reminiscent of boundary-layer solutions in the dynamics of viscous fluids at high Reynolds number. However, we caution the reader that in our problem the highest derivatives in the equations are important everywhere when $p\gg1$, not just in the boundary layers.} approaches a constant tilt, $\beta\to\beta_\text{p}+C_\text{in}(1+\gamma_\text{in})$, at large $x$. In the same limit, the dimensionless torque approaches a constant value, $f\to C_\text{in}(1-\gamma_\text{in})$, which is smaller than the relative tilt $\beta-\beta_\text{p}$, but only slightly so if the inner boundary is well into the boundary-layer solution so that $\gamma_\text{in}\ll1$. The outer approximate solution approaches a constant tilt, $\beta\to\beta_\text{s}-C_\text{out}(1+\gamma_\text{out})$, at small $x$. In the same limit, the dimensionless torque approaches a constant value, $f\to C_\text{out}(1-\gamma_\text{out})$, which is smaller than the relative tilt $\beta_\text{s}-\beta$, but only slightly so if the outer boundary is well into the boundary-layer solution so that $\gamma_\text{out}\ll1$.

If $p$ is sufficiently large ($p\gtrsim50$, say), then there is a well-defined intermediate region in which $\beta\approx\text{constant}$ and $f\approx\text{constant}$. In this case, matching $\beta$ and $f$ in the intermediate region gives
\begin{equation}
  C_\text{in}=\left(\f{1-\gamma_\text{out}}{1-\gamma_\text{in}\gamma_\text{out}}\right)\f{\beta_\text{s}-\beta_\text{p}}{2},\qquad
  C_\text{out}=\left(\f{1-\gamma_\text{in}}{1-\gamma_\text{in}\gamma_\text{out}}\right)\f{\beta_\text{s}-\beta_\text{p}}{2}, \label{Cinout}
\end{equation}
which are almost equal if $\gamma_\text{in}\ll1$ and $\gamma_\text{out}\ll1$, in which case the intermediate tilt is the mean of the inner and outer tilts (which differs from $\beta_0$ given by Equation (\ref{beta0})).

In Appendix \ref{sec:comp} we compare the results
of the expansion method in Section \ref{sec:warm} to
the boundary layer method in this section.  Both methods produce similar results in the limit
of large $p$ at  radii where the disc warp is weak.

The inner and outer solutions can be combined to provide
a global solution given by
\begin{equation}
\beta(x)=
\begin{cases}
  \beta_{\rm in}(x) & \text{ for } x \le 1, \\
 \beta_{\rm out}(x) & \text{ for } x > 1, 
\end{cases} \label{betainout}
\end{equation}
where $\beta_{\rm in}$ and $\beta_{\rm out}$ are the $\beta$
functions given by Equations (\ref{betain}) and (\ref{betaout}), respectively.
This solution provides accurate results for $\beta$
but has a small jump at $x=1$.

For the case of a cool disc with $\gamma_\text{in}\ll 1$ and  $\gamma_\text{out}\ll 1$, we see from Equation (\ref{betain}) that for  
$x_{\rm in} < x \ll $1  
the disc tilt with respect to the planet's equator
 varies as $\exp{(-1/(2 p x^2))}$. Consequently, we adopt an effective disc inner Laplace radius for a gaseous disc given by
\begin{equation}
x_{\rm L, in}= \frac{1}{\sqrt{p}}, \label{xLin}
\end{equation}
corresponding to dimensional radius
\begin{equation}
r_{\rm L, in} =  \sqrt{\frac{ 3J_2}{h}} R_{\rm p} = 2^{1/5} \frac{r_{\rm L}}{\sqrt{p}},
\end{equation}
where $r_{\rm L}$ is the classical Laplace radius for particles.
These results are consistent with the estimate in Equation (\ref{xilim})  for 
the range of validity of the disc inner radius for the weak warp approximation in Section \ref{sec:warm}. The disc cannot
extend to small enough radii to contain a gas disc inner Laplace radius because typically
$r_{\rm L, in} < R_{\rm p}$.
This Laplace radius is small compared compared to the classical Laplace radius for
  typical parameters that we have argued involves $p \gg 1$.
 For the disc to be close to coplanar with the equator of the planet, within about 10\%
 of the planet obliquity, requires an even smaller disc inner radius 
 $r_{\rm in} \lesssim 0.5  r_{\rm L, in}$.
 
 Similarly, we see from Equation (\ref{betaout}) that for  $\gamma_\text{in}\ll 1$,  $\gamma_\text{out}\ll1$, and
$x_{\rm out} > x \gg $1, 
the disc tilt with respect to the planet's orbital plane varies as $\exp{(- x^3/(3 p))}$. We adopt an effective disc outer Laplace radius for a gaseous disc given by
\begin{equation}
x_{\rm L, out}= p^{1/3}, \label{xLout}
\end{equation}
corresponding to dimensional radius
\begin{equation}
r_{\rm L, out} =  (2 h)^{1/3} R_{\rm H},
\end{equation}
where $R_{\rm H}$ is the Hill radius of the planet.
These results are consistent with the estimate in Equation (\ref{xolim})  for 
the range of validity of the disc outer radius for the weak warp approximation in Section \ref{sec:warm}.

A cool disc is in good radial communication, $p\gg1$, and so is flat at intermediate radii $x\sim 1$ ($r \sim r_{\rm L}$).
A cool disc undergoes warping in its inner and outer regions because $h$ is sufficiently small that
$r_{\rm in} < (3 J_2/h)^{1/2} R_{\rm p}$ and 
$r_{\rm out} > (2 h)^{1/3} R_{\rm H}$, respectively.

We can also express $p$ as
\begin{equation}
p^{5/6} = \left( \frac{ x_{\rm  in } }{ x_{\rm L, in} } \right) \left( \frac{x_{\rm L, out} }{x_{\rm out} } \right) \frac{x_{\rm out} }{ x_{\rm in }}  = \left( \frac{ r_{\rm  in } }{ r_{\rm L, in} } \right) \left( \frac{r_{\rm L, out} }{r_{\rm out} } \right) \frac{r_{\rm out} }{ r_{\rm in }}. \label{pxinxout}
\end{equation}
The terms in parentheses are of order unity for the warm disc (Model B) plotted in Figure \ref{fig:betastd}, while $p$ is large because 
$x_{\rm out} \gg x_{\rm in}$. For that model $x_{\rm in} \sim 1.5 x_{\rm L,in}$ and $x_{\rm out} \sim 0.7 x_{\rm L,out}$. The warm disc has $p\gg 1$ and does not extend to either the inner or outer Laplace radius, which explains why it undergoes only weak warping.

\begin{figure}
	\includegraphics[width=0.8\textwidth]{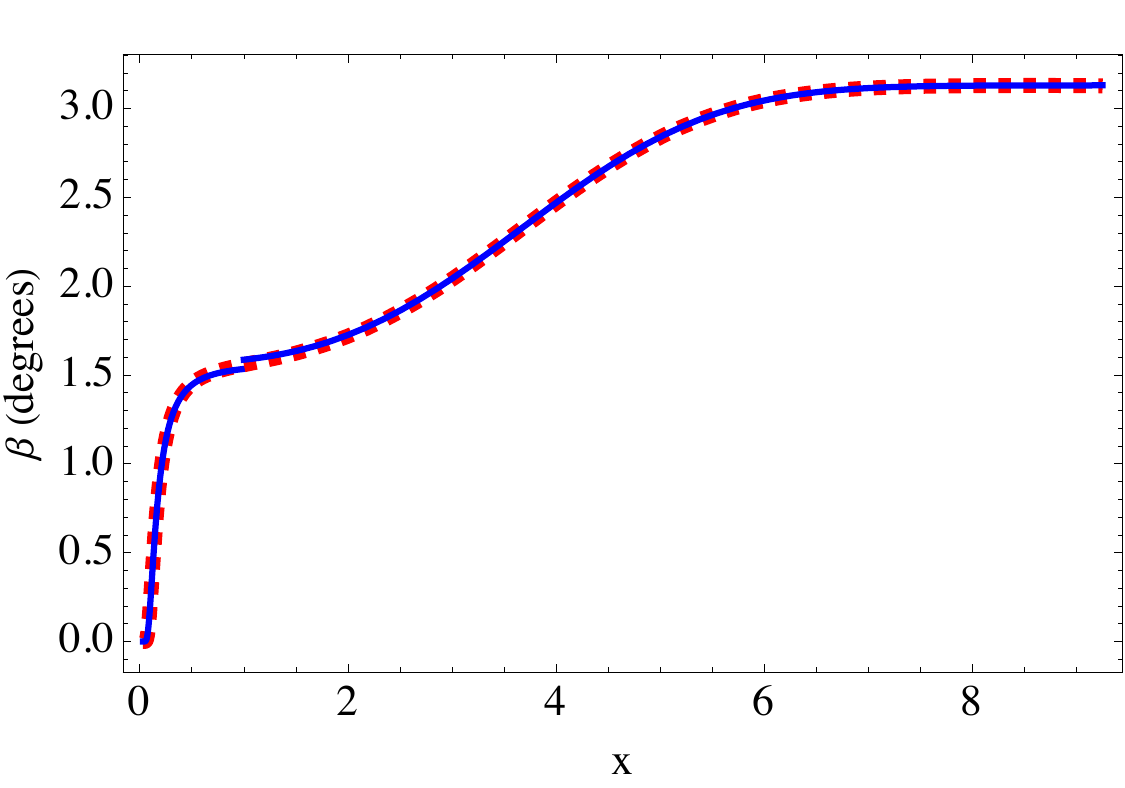}\\
    	\includegraphics[width=0.8\textwidth]{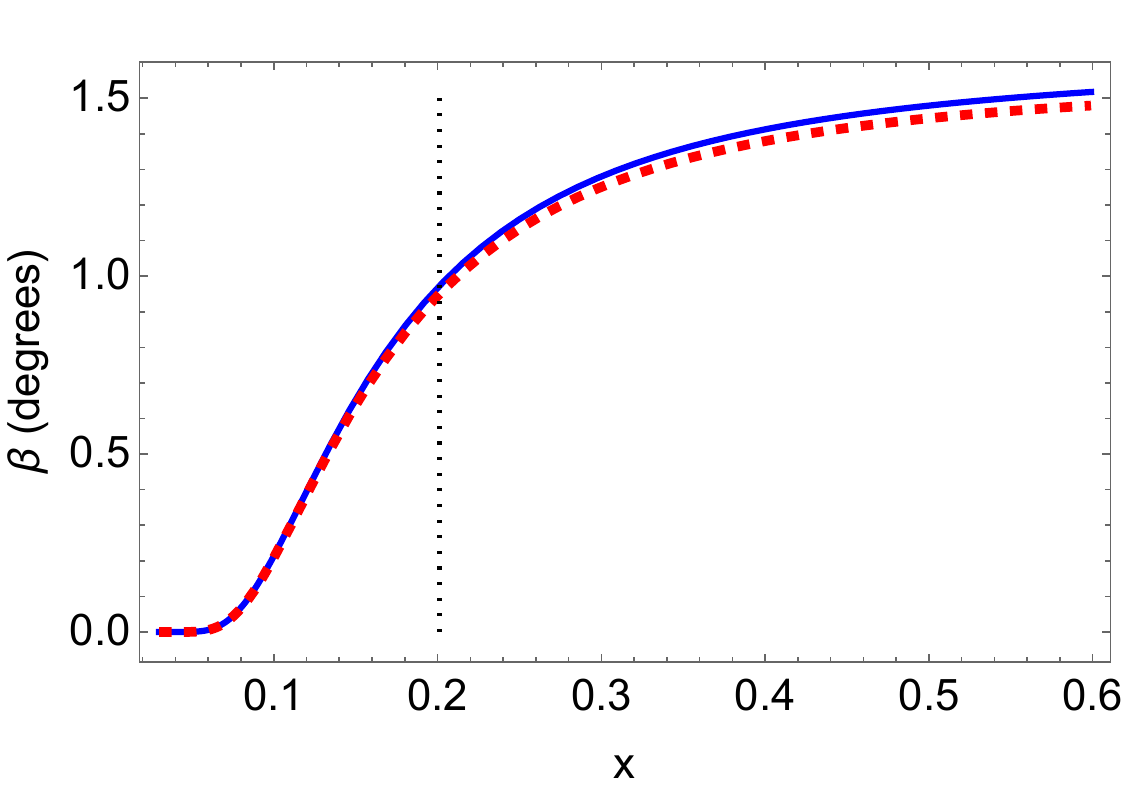}\\
    \caption{Plot of $\beta(x)$, the local tilt angle of the Laplace surface relative to the equatorial plane of Jupiter, as a function of dimensionless distance from the centre of Jupiter
   for Model D  that has aspect ratio $h=0.001$.
   The blue line is based on a numerical solution
    to Equations (\ref{beta})--(\ref{BCf}). 
    The overlapping dashed red line is obtained from the  analytic expressions in Equations (\ref{betainout}) with $\beta_{\rm p}=0$. 
    The lower panel is a closeup of the upper panel. The dotted vertical line denotes the disc inner Laplace radius given by 
    Equation (\ref{xLin}). }    
    \label{fig:betacool}
\end{figure}

\begin{figure}
   \includegraphics[width=0.8\textwidth]{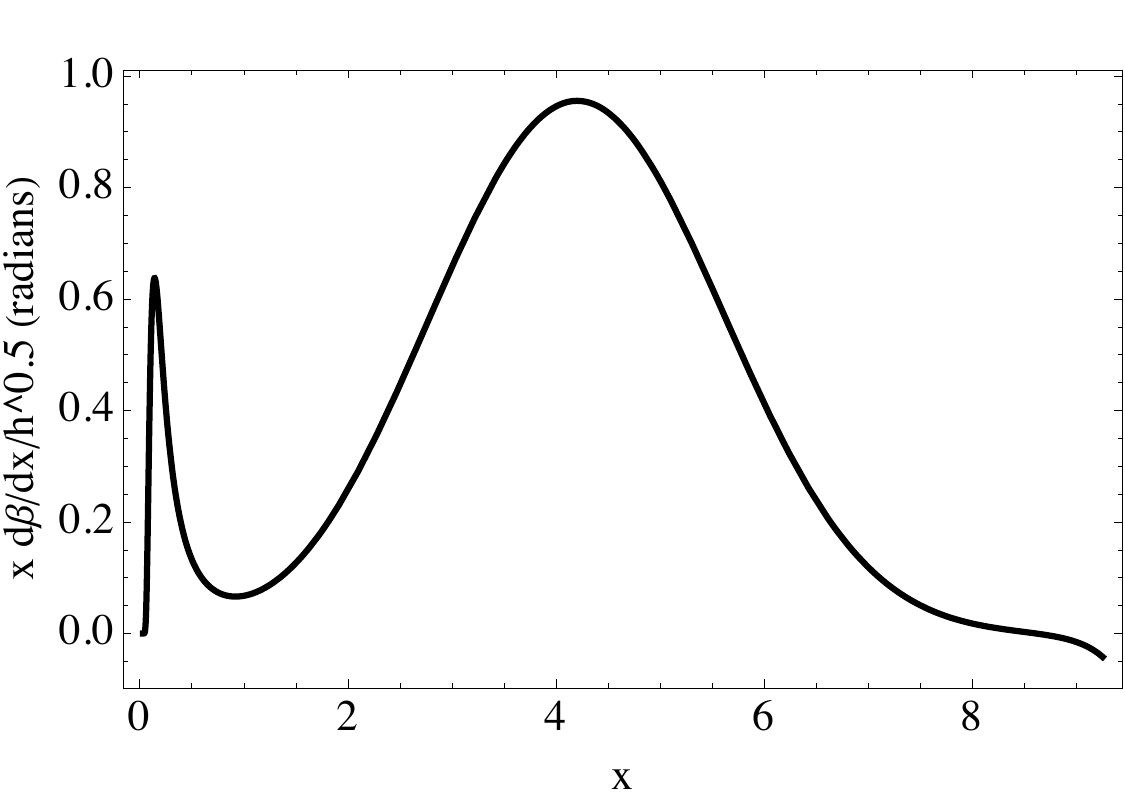}
    \caption{Plot of the scaled warp, $x \, \dd \beta/\dd x/\sqrt{h}$, as a function of dimensionless distance from the centre of Jupiter for Model D that is plotted in Figure~\ref{fig:betacool}. }    
    \label{fig:warpcool}
\end{figure}

We consider a case for which a disc is cool enough to contain the inner and outer disc Laplace radii.
We calculate Model D that has the same parameters as Model B plotted in 
Figure~\ref{fig:betastd} but with $h=0.001$. In this case, $p=24.8,  x_{\rm in} \approx 0.14x_{\rm L, in}$, and
$x_{\rm out} \approx 3.2 x_{\rm L, out}$. The tilt of 
of the disc relative to the equator of the planet is shown in Figure~\ref{fig:betacool}. The numerically determined disc
tilt agrees well with the analytic model given by Equation (\ref{betainout}).
  For $x \lesssim 0.05 < x_{\rm L,in} \approx 0.2$
the tilt is very small.
Figure \ref{fig:warpcool} plots $x \dd\beta/\dd x/\sqrt{h}$ as a function of $x$ which is a measure of the degree of nonlinearity
\citep{Ogilvie2006}.
The two peaks occur near the two Laplace radii and indicate that nonlinearities may play a role in the warp structure.

\section{Effects of disc self-gravity}
\label{sec:sg}

We consider here the effects of disc self-gravity on the shape of the Laplace surface.
The calculation of the precession rate due to self-gravity needs to be done carefully to avoid spurious
effects due to singularities \citep{Sefilian2019}.
Equations (\ref{beta}) and (\ref{f}) are modified to give
\begin{equation}
\frac{\beta - \beta_{\rm p}}{x^3} + (\beta-\beta_{\rm s}) x^2  + p  \lambda T_{\rm sg} = p \frac{\dd f}{\dd x} \label{betasg}
\end{equation}
and
\begin{equation}
\left(\frac{1}{x^3} + x^2  + p \lambda \omega_{\rm sg} \right) f = p \frac{\dd \beta}{\dd x} \label{fsg},
\end{equation}
where
\begin{equation}
\lambda = \frac{M_{\rm d}}{2 h x_{\rm out} M_{\rm p}} = \frac{1}{ Q_{\rm out} \, x_{\rm out} },
\end{equation}
 $Q_{\rm out}$ is the Toomre $Q$ parameter at $x=x_{\rm out}$ that is the lowest $Q$ value in the disc,
\begin{equation}
T_{\rm sg}(x)  = \int_{x_{\rm in}}^{x_{\rm out}} \left(\frac{x_<}{x^2_>}\right) b^{(1)}_{3/2}\left( \frac{x_<}{x_>} \right)
(\beta(x) - \beta(x')) \, \dd x', \label{Tsg}
\end{equation}
and
\begin{equation}
\omega_{\rm sg}(x)  = \frac{\dd}{\dd x} \left[ x^2 \frac{\dd}{\dd x} \int_{x_{\rm in}}^{x_{\rm out}} b^{(0)}_{1/2}\left(\frac{x_<}{x_>}\right) \frac{\dd x'}{x'} \right], \label{omsg}
\end{equation}
where $x_< = \text{min}(x,x')$ and  $x_> = \text{max}(x,x')$.
$T_{\rm sg}$ is singular but provides finite results when softened by an amount that depends on disc aspect ratio $h$.
The softening is performed by replacing
$b^{(1)}_{3/2}(x)$
by 
\begin{equation}
B^{(1)}_{3/2}(x) = \frac{2}{\pi} \int_0^\pi 
\frac{\cos u \, \dd u}{(1-2 x \cos u + x^2 +h^2)^{3/2}}.
\end{equation}
With the softened 
Laplace coefficient $B^{(1)}_{3/2}$, torque $T_{\rm sg}(x)$ is nonsingular and converges in the limit of small $h$.
Near the disc edges, the apsidal precession rate $\omega_{\rm sg}$ undergoes rapid changes. Consequently, we apply softening by replacing
$b^{(0)}_{1/2}(x)$
with
\begin{equation}
B^{(0)}_{1/2}(x) = \frac{2}{\pi} \int_0^\pi 
\frac{\dd u}{(1-2 x \cos u + x^2 +h^2)^{1/2}}.
\end{equation}


The self-gravity term in Equation (\ref{betasg}) complicates the analysis because the equations become integro-differential equations. However, that
self-gravity term is small compared to the other terms in that equation for larger  $Q_{\rm out}$. It can also be small because it vanishes for a rigid tilt. We use an iteration scheme to solve Equations (\ref{betasg}) and (\ref{fsg}). In the first step,
the self-gravity term in Equation (\ref{betasg}) is ignored, as would apply for a rigid tilt. The equations
are solved as differential equations. Using the solution for $\beta(x)$,
we evaluate the self-gravity term in Equation (\ref{betasg}) and then use it to 
solve Equations (\ref{betasg}) and (\ref{fsg}). We repeat this process multiple times until the maximum change in $\beta(x)$ between successive iterations is less than 0.1\%. Convergence is slower for lower $Q_{\rm out}$ values. For $Q_{\rm out} \ge 1$, the scheme converges in less than 4 iterations. 

\begin{figure}
	\includegraphics[width=0.8\textwidth]{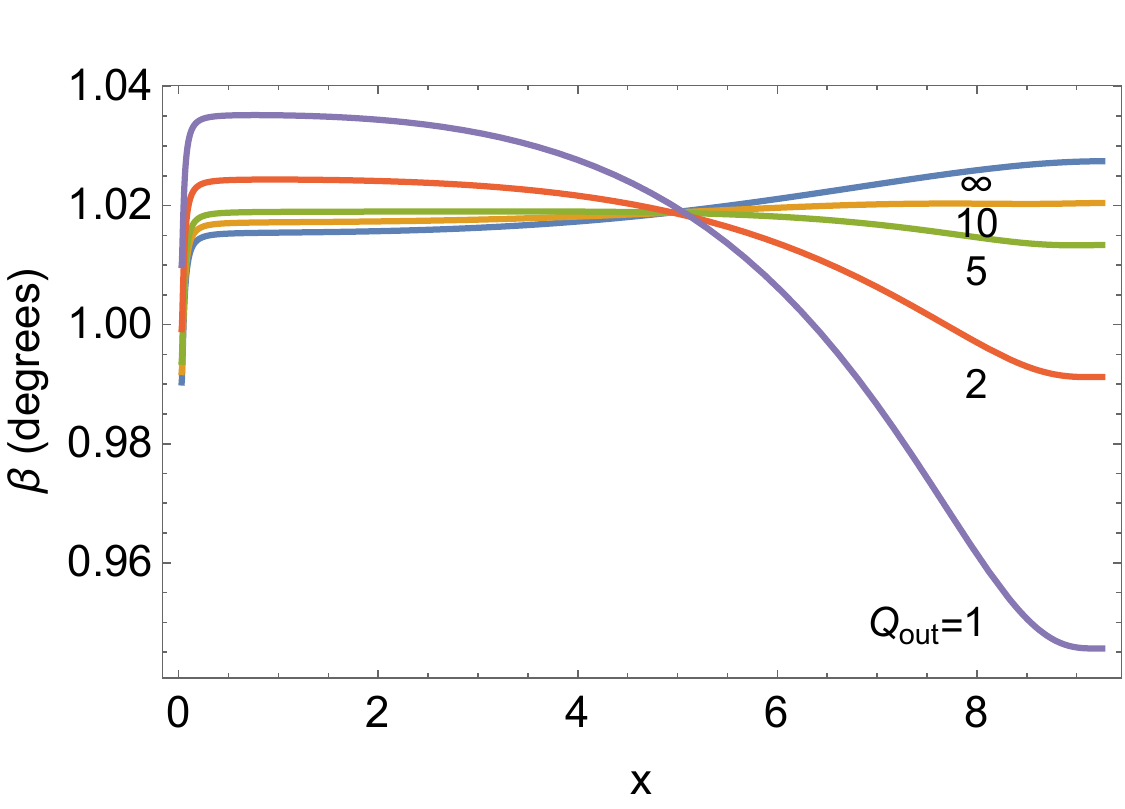}\\
    \caption{Plot of $\beta(x)$, the tilt angle of the Laplace surface, as a function of dimensionless distance from the centre of Jupiter for Model E with different values of the self-gravity  parameter $Q_{\rm out}$. }
    \label{fig:betasg}
\end{figure}

Figure~\ref{fig:betasg} plots the resulting Laplace surface profiles for Model E with various values of $Q_{\rm out}$.  The case of infinite $Q_{\rm out}$
matches the blue line in  Figure \ref{fig:betastd}.  Notice that due to self-gravity, the disc warp is reduced for $Q_{\rm out}=10$, but increases somewhat  for  $Q_{\rm out} \le 2$. 
The effects of self-gravity cancel near $x\simeq 5.1$ resulting in nearly the same value of $\beta$ for all $Q_{\rm out}$. 
 \cite{Ward1981} approximately solves Equation (\ref{betasg}) under the assumption that $f(x)=0$.
The effects of self-gravity in  Figure~\ref{fig:betasg} bear some resemblance
to the pressureless case
 seen in Figure~5 of \cite{Ward1981}, in which
there is also a radius where the effects of self-gravity cancel.
In that case the tilt at large radii decreases with higher disc mass, as
 we also find.

The disc tilt is not significantly changed due to the effects of self-gravity. The disc inner edge
remains misaligned with  the equator of the planet. The level of misalignment at the inner edge increases slightly with decreasing $Q_{\rm out}$ 
(increasing disc mass).

\section{Effects of disc viscosity}
\label{sec:visc}

\subsection{Steady solutions}
\label{s:steady_viscous}

We consider here the effects of a non-zero viscosity $\alpha$ parameter in Equation (\ref{dFdt}).
The viscosity causes a phase lag in the disc tilt that results in a tilt component in the $y$-direction.
 The torque equations for the circumplanetary disc with viscosity are given by Equations (\ref{dldt}) and (\ref{dFdt}).
It is convenient to encode the vectors as complex quantities such that any vector 
$\bm{v}$ is expressed as $v= v_x + \ii v_y$. 
For complex tilt $\ell(r,t) = \ell_x(r,t) + \ii \ell_y(r,t)$.
 In complex form, torque equations for the circumplanetary disc 
are given by
\begin{align}
&\Sigma r^2 \Omega \frac{\partial \ell}{\partial t} =\frac{1}{r}\frac{\partial F}{\partial r}+ \Sigma T \label{dldtc}\\
& \text{and} \nonumber \\
& \frac{\partial F}{\partial t} - \ii \omega_{\rm a}  F +\alpha \Omega  F =\frac{1}{4} \Sigma H^2 r^3 \Omega^3 \frac{\partial \ell}{\partial r}. \label{dFdtc}
\end{align}

 In this section, we denote a tilt by $\beta$ following our previous notation, so that $\beta = \ell = \ell_x + \ii \ell_y$.
 We again consider the case that $ \partial \ell/\partial t=0, \partial F/\partial t=0, \Sigma(r) = \Sigma_0/r$ with constant $\Sigma_0$ and disc aspect ratio $h=H/r$ constant.
Under these simplifying assumptions, Equations (\ref{dldtc}) and (\ref{dFdtc}) are respectively written in dimensionless form as
\begin{equation}
\frac{\beta - \beta_{\rm p}}{x^3} + (\beta-\beta_{\rm s}) x^2 = p \frac{\dd f}{\dd x} \label{betavisc}
\end{equation}
and
\begin{equation}
\left(\frac{1}{x^3} + x^2 \right) f  + \frac{2 \ii \alpha p f }{h x}= p \frac{\dd \beta}{\dd x} \label{fvisc},
\end{equation}
where
$x$ and $p$ are defined as before in Equations (\ref{xdef}) and (\ref{pdef}) respectively and
 \begin{equation}
  f = \frac{- 2 \ii F}{\Sigma_0 h G M_{\rm p}}.
\end{equation}
We again apply as boundary conditions that $f$ vanishes at the disc boundaries, as given by Equation (\ref{BCf}).

Equations (\ref{betavisc}) and (\ref{fvisc}) reduce to Equations (\ref{beta}) and (\ref{f}) in the case
of $\alpha=0$. Owing to viscosity, there is an additional term in Equation (\ref{fvisc}).
Appendix \ref{sec:betaviscanalytic} describes an analytic solution for the complex tilt of a warm disc by using a
perturbation expansion in $\epsilon=1/p$ that extends the approach taken in Section \ref{sec:warm}.
The results show that the change $\delta \beta$ in tilt due to the viscosity scales with $\alpha$ and $h$
 as
\begin{align}
\text{Re}(\delta \beta)  \propto (\beta_{\rm s} - \beta_{\rm p})  \left( \frac{\alpha}{h} \right)^2\quad \text{and} \qquad 
\text{Im}(\delta \beta)  \propto  (\beta_{\rm s} - \beta_{\rm p})  \left( \frac{\alpha}{h} \right).
\end{align}
The  constants of proportionality depend on $x, x_{\rm in}$, $x_{\rm out}$, and $p$.
They are typically somewhat less than unity for the warm disc models we consider.
 In the wavelike regime, we have that $\alpha < h$ and consequently the effects of viscosity
 on the tilt are not large and are small for $\alpha \ll h$.

\begin{figure}
	\includegraphics[width=0.8\textwidth]{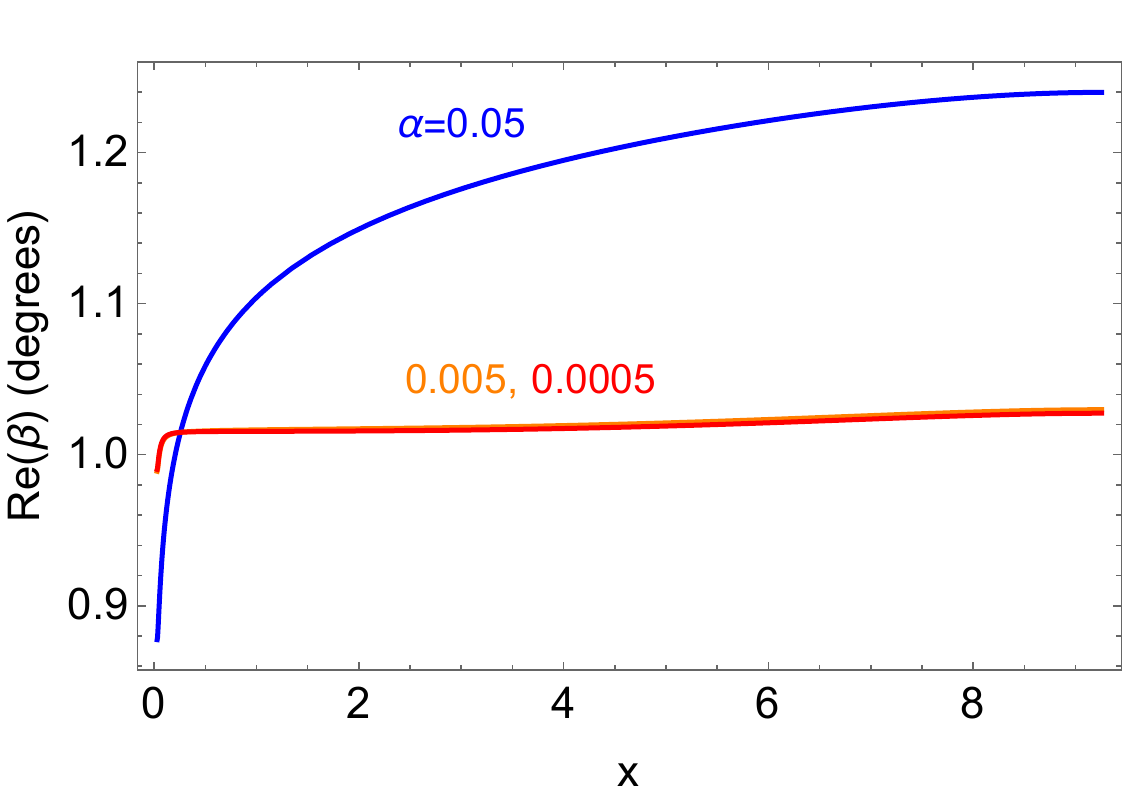}\\
    \includegraphics[width=0.8\textwidth]{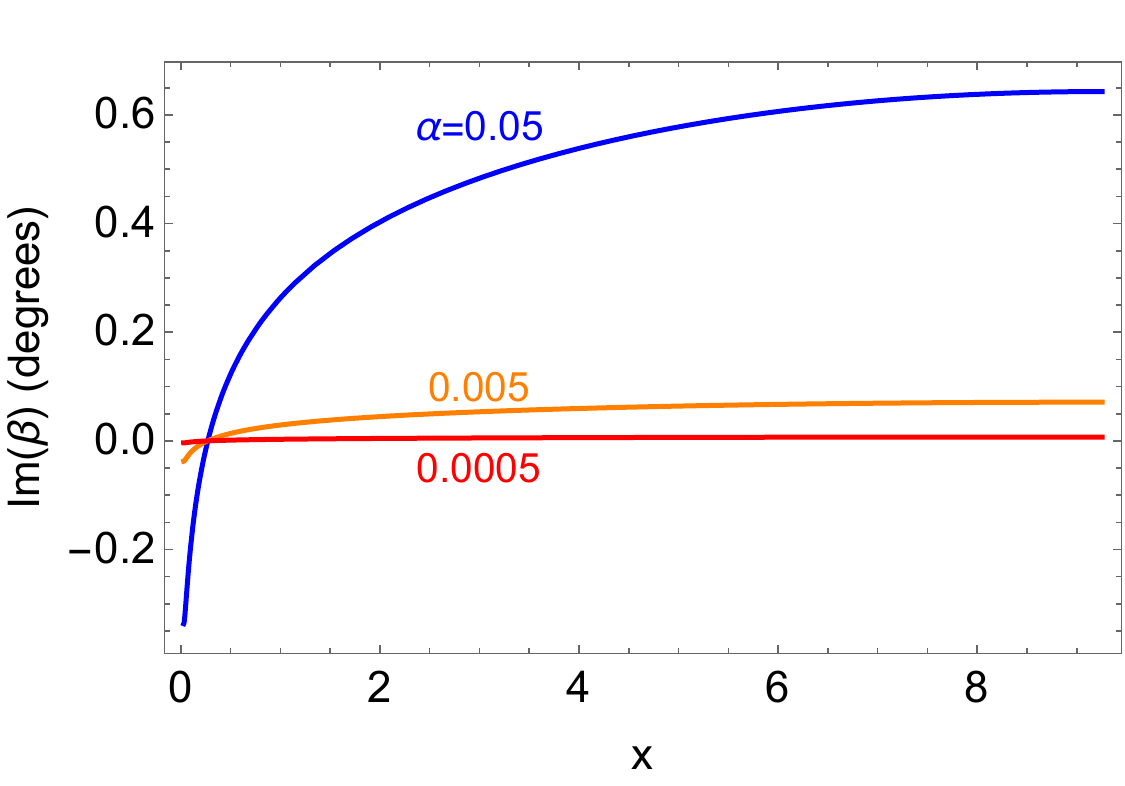}
    \caption{Plot of the real and imaginary parts $\beta(x)$, the local tilt angle of the Laplace surface in the $x-z$ and $y-z$ planes, respectively, for different values of the viscosity parameter $\alpha$ in Model F.    The curves for $\alpha =0.0005$ and 0.005 in the upper plot nearly overlap.}
    \label{fig:betastdvisc}
\end{figure}

Figure~\ref{fig:betastdvisc} plots the real and imaginary parts of tilt angle $\beta$ relative to the equator of Jupiter as a function of  dimensionless distance $x$ from the planet for Model F, the warm disc the model plotted in Figure~\ref{fig:betastd} but with viscosity. The plot is based on numerical solutions 
to Equations (\ref{betavisc}) and (\ref{fvisc}) with boundary conditions given by Equation (\ref{BCf}) . The effect of viscosity is to increase the warp for both the real and imaginary
parts of $\beta$.  But for all three models, the degree of nonlinearity is small: $(x \, \dd \, \text{Re}(\beta)/\dd x) /\sqrt{h} < 0.015$ and $(x \, \dd \, \text{Im}(\beta)/\dd x) /\sqrt{h} < 0.005$.
The case of $\alpha=0.05 = h/2$ shows the largest change due to viscosity in the plot. But even in this case, the disc does
not align the the equator of the planet near its inner edge. For the plotted smaller values
of $\alpha = h/20$ and $h/200$, the effects of viscosity on the disc tilt are quite small.
Consequently,  the pressure effects dominate the disc shape for $\alpha \ll h$, 
resulting in a highly flattened disc.

\subsection{Evolution towards the steady state}
\label{sec:viscevol}

We consider the linear partial differential equations (PDEs) given by Equations (\ref{dldtc}) and (\ref{dFdtc})  for $\ell(r,t)$ and $F(r,t)$ and include the effects of self-gravity on the torque and apsidal precession rate
given by Equations (\ref{Tsg}) and (\ref{omsg}), in additional to gravitational effects of the star and planet.
Equations (\ref{dldtc}) and (\ref{dFdtc}) are forced linearly by the inner and outer tilts $\beta_\text{p}$ and $\beta_\text{s}$. (By subtracting either $\beta_\text{p}$ or $\beta_\text{s}$ from $\ell$, the problem can be reformulated in such a way that it is forced only by the misalignment $\beta_\text{s}-\beta_\text{p}$.)


The general solution of these equations   can be regarded as a sum of the particular, steady solution described in Section~\ref{s:steady_viscous} and a complementary function, which is the general solution of the same PDEs in which the forcing terms $\beta_\text{p}$ and $\beta_\text{s}$ are set to zero.  The complementary function describes a departure from the Laplace surface.
Physically, this complementary function represents an arbitrary, time-dependent, free warp in a system with no misalignment. In the presence of self-gravity and viscosity, such an unforced warp must decay to zero. This can be shown from the evolution of the angular-momentum deficit (AMD):
\begin{equation}
  \f{\dd}{\dd t}\int_{r_\text{in}}^{r_\text{out}}\Sigma r^2\Omega\left(\f{1}{2}|\ell|^2+\f{1}{2}|f|^2\right)\,2\pi r\,\dd r=-\int_{r_\text{in}}^{r_\text{out}}\Sigma r^2\Omega\left(\alpha\Omega|f|^2\right)\,2\pi r\,\dd r,
\end{equation}
which follows from Equations (\ref{dldtc}) and (\ref{dFdtc}) after an integration by parts and the use of the boundary conditions. Since the right-hand side is negative definite, the (relative) AMD must decay until $f=0$ throughout the disc. The precessional terms do not contribute directly to this argument, although they do ensure that $f=0$ also implies $\ell=0$. The complementary function decays to zero. 
Consequently, the Laplace surface is the final state of a self-gravitating, viscous  disc for any initial condition, provided that linear theory holds throughout the evolution.
 In particular, the steady solution described in Section~\ref{s:steady_viscous} is the final state of the linear evolution of a non-self-gravitating, viscous  disc.

\section{Discussion and Summary}
\label{sec:summary}

We determined the shape of the Laplace surface of a circumplanetary
disc that results from accretion of circumstellar gas involving a planet with small obliquity, such as Jupiter.
We applied the linear theory of warped discs that includes the effects of pressure, self-gravity, and viscosity for a disc in the wavelike regime in which
$\alpha \ll h$, where $h=H/r$ is the aspect ratio of the disc. Owing to the effects of gas pressure, the disc is highly flattened (unwarped) and differs considerably from the classical Laplace surface 
that is based on particles. In typically warm discs, self-gravity and viscosity can act to enhance the disc warp,
but their effects are typically small.  Due to dissipation, a disc that initially does
not lie on its Laplace surface will evolve to it, as shown
for small initial departures from the Laplace surface in Section 
\ref{sec:viscevol}.

The response of the disc to the gravitational torques due to the planet and central
star is controlled by parameter $p$ defined in Equation (\ref{pdef}).
It measures the level of radial communication provided by gas pressure
that acts to flatten the disc against the effects of the two mutually misaligned gravitational quadrupoles that act to warp the disc. The parameter is large,  $p \simeq 2.5 \times 10^3$, for a circumjovian disc with $h =0.1$, and is proportional to $h$.

There are three regimes of disc Laplace surfaces.
In the cold disc case, $p \ll 1$ and the disc is accurately described by a model based 
on ballistic particles. Close to the planet, the Laplace surface lies along the planet's equator and far from the planet 
it lies along the planet's orbital plane.
The Laplace surface undergoes a warp 
at its Laplace radius $r_{\rm L} \simeq 29 R_{\rm J}$ in the case of Jupiter. But this case requires the disc aspect ratio to be very small,
$h < 10^{-4}$. Such values are well outside the expected range for circumplanetary discs, which are expected to have $h\sim 0.1$ or greater \citep[e.g.,][]{Batygin2020, Martin2023}.
In the cool disc case,  the disc is in good overall radial communications because $p \gg 1$.
The disc is flat at intermediate radii.
But the disc undergoes significant warping in its inner and outer regions where tidal torques are strongest.  There are in effect two Laplace radii $r_{\rm L, in} =  2^{1/5} r_{\rm L}/\sqrt{p}$ and $r_{\rm L, out} = (2 h)^{1/3} R_{\rm H}$, for planet Hill radius $R_{\rm H}$.
Interior to the inner Laplace radius, the disc aligns with the equator of the planet (see Figure~\ref{fig:betacool}). Exterior to the 
outer Laplace radius, the disc aligns with the orbital plane of the planet.
For a cool disc the inner and outer radii respectively satisfy $r_{\rm in} < r_{\rm L, in}$ and $r_{\rm out} > r_{\rm L, out}.$
The cool disc case occurs for $h$ values that are larger than the cold disc case (e.g., $h\sim 0.001$),
but are also
unrealistically small for a circumjovian disc. In addition, the disc may be subject to nonlinear effects near the Laplace radii
(see Figure~\ref{fig:warpcool}).
In the warm disc case, $p \gg 1$
and the disc does not cover its inner or outer Laplace radii, $r_{\rm in} > r_{\rm L, in}$ and $r_{\rm out} < r_{\rm L, out}$.
This case occurs for larger 
disc aspect ratios $h \gtrsim 0.1$ that are expected for a circumjovian disc.
In this case, the disc is highly flattened (i.e., unwarpred) with weak warping at the inner
and outer disc edges (see Figure~\ref{fig:betastd}). The warp is sufficiently
weak to be in the linear regime  (see Figure~\ref{fig:warpstd}).
The inner edge of the disc does not align with the equator
of the planet. The disc tilt is intermediate between the tilt of the planet
and the tilt of the planet's orbital plane (see Figure~\ref{fig:beta0rin}).

The analysis in this paper made some simplifying assumptions.
We considered models with $\Sigma \propto 1/r$ and constant $H/r$.
Such models are plausible, but other density and aspect ratios should
be considered. The precessional torques we applied may not be accurate
at larger orbital radii in the outer parts of the Hill sphere.
For larger values of disc aspect ratios $h > 0.2$, non-Keplerian effects
could play a role \citep{Martin2023}. 
The disc tilt is subject to a global instability
\citep{Lubow2000, Martin2020, Martin2021a}.
Accretion from the circumstellar disc onto the circumplanetary disc
  provides an additional torque that acts to align the disc with the orbital plane of the planet that is not taken into account. This torque would act to align the disc towards to orbital plane of the planet.
  We have applied a zero-angular-momentum-flux inner boundary condition on the disc.
Torques due to the planetary magnetic field could affect that boundary condition.
We adopted the current value for the dimensionless quadrupole moment of Jupiter. 
It could have been different at early times and may have been smaller due to magnetic braking \citep{Batygin2018}.
A smaller value of the quadrupole moment would increase the level of disc-planet misalignment near the inner edge of the disc.
However, even taking into account these issues, we expect
that the main qualitative effect of disc flattening 
 will still hold for a circumjovian disc, since parameter $p$ is very 
 large, indicating that the disc is subject to strong radial communication by pressure.

 The linear theory of warps that we apply is justified for a circumjovian disc because of the small obliquity of Jupiter. In the case of Saturn with its large obliquity of $26.7^{\circ}$, it is not clear how well our results carry over. Nonetheless, we expect that its circumplanetary disc experienced large departures from its classical Laplace surface. Such departures might explain the large orbital tilt of Iapetus away from its classical Laplace surface. On the other hand,
the orbit of Titan, with an orbital radius that is smaller than that of Iapetus, lies close to the classical  Laplace surface and is inclined to the orbit of Iapetus. It is not clear that the both tilts can be explained by formation within a misaligned disc, especially if it is nearly flat.

 The regular satellites of Jupiter lie somewhat close the the classical
 Laplace surface (see \url{https://ssd.jpl.nasa.gov/sats/elem/}).  Of the eight moons closest to Jupiter, three lie within than $0.1^\circ$ of the Laplace surface. 
The others are inclined by $0.2^\circ$ to $1.1^\circ$ from the classical
 Laplace surface. It has been thought that the circumplanetary disc
 coincided with the classical Laplace surface which was the site of satellite formation \citep[e.g.,][]{Tremaine2009}. Interactions between moons may have led to departures
 from the Laplace surface.
 The results here suggests that they formed from a disc that was 
 inclined by $\sim 1^\circ$ from the equator of Jupiter.
 The moons may have  evolved closer to the classical Laplace surface
 during or after their formation in the circumjovian disc.

\section*{Acknowledgements}
We made use of Mathematica software in our derivations.
 GIO acknowledges support from STFC through grant ST/X001113/1.

\section*{Data Availability}
The data underlying this article will be shared on reasonable request to the corresponding author.



\bibliographystyle{mnras}
\bibliography{main} 

\begin{thebibliography}{}
\makeatletter
\relax
\def\mn@urlcharsother{\let\do\@makeother \do\$\do\&\do\#\do\^\do\_\do\%\do\~}
\def\mn@doi{\begingroup\mn@urlcharsother \@ifnextchar [ {\mn@doi@}
  {\mn@doi@[]}}
\def\mn@doi@[#1]#2{\def\@tempa{#1}\ifx\@tempa\@empty \href
  {http://dx.doi.org/#2} {doi:#2}\else \href {http://dx.doi.org/#2} {#1}\fi
  \endgroup}
\def\mn@eprint#1#2{\mn@eprint@#1:#2::\@nil}
\def\mn@eprint@arXiv#1{\href {http://arxiv.org/abs/#1} {{\tt arXiv:#1}}}
\def\mn@eprint@dblp#1{\href {http://dblp.uni-trier.de/rec/bibtex/#1.xml}
  {dblp:#1}}
\def\mn@eprint@#1:#2:#3:#4\@nil{\def\@tempa {#1}\def\@tempb {#2}\def\@tempc
  {#3}\ifx \@tempc \@empty \let \@tempc \@tempb \let \@tempb \@tempa \fi \ifx
  \@tempb \@empty \def\@tempb {arXiv}\fi \@ifundefined
  {mn@eprint@\@tempb}{\@tempb:\@tempc}{\expandafter \expandafter \csname
  mn@eprint@\@tempb\endcsname \expandafter{\@tempc}}}

\bibitem[\protect\citeauthoryear{{Artymowicz} \& {Lubow}}{{Artymowicz} \&
  {Lubow}}{1996}]{Artymowicz1996}
{Artymowicz} P.,  {Lubow} S.~H.,  1996, \mn@doi [\apjl] {10.1086/310200}, \href
  {https://ui.adsabs.harvard.edu/abs/1996ApJ...467L..77A} {467, L77}

\bibitem[\protect\citeauthoryear{{Ayliffe} \& {Bate}}{{Ayliffe} \&
  {Bate}}{2009}]{Ayliffe2009}
{Ayliffe} B.~A.,  {Bate} M.~R.,  2009, \mn@doi [\mnras]
  {10.1111/j.1365-2966.2009.15002.x}, \href
  {https://ui.adsabs.harvard.edu/abs/2009MNRAS.397..657A} {397, 657}

\bibitem[\protect\citeauthoryear{{Batygin}}{{Batygin}}{2018}]{Batygin2018}
{Batygin} K.,  2018, \mn@doi [\aj] {10.3847/1538-3881/aab54e}, \href
  {https://ui.adsabs.harvard.edu/abs/2018AJ....155..178B} {155, 178}

\bibitem[\protect\citeauthoryear{{Batygin} \& {Morbidelli}}{{Batygin} \&
  {Morbidelli}}{2020}]{Batygin2020}
{Batygin} K.,  {Morbidelli} A.,  2020, \mn@doi [\apj]
  {10.3847/1538-4357/ab8937}, \href
  {https://ui.adsabs.harvard.edu/abs/2020ApJ...894..143B} {894, 143}

\bibitem[\protect\citeauthoryear{{Benisty} et~al.,}{{Benisty}
  et~al.}{2021}]{Benisty2021}
{Benisty} M.,  et~al., 2021, \mn@doi [\apjl] {10.3847/2041-8213/ac0f83}, \href
  {https://ui.adsabs.harvard.edu/abs/2021ApJ...916L...2B} {916, L2}

\bibitem[\protect\citeauthoryear{{Canup} \& {Ward}}{{Canup} \&
  {Ward}}{2002}]{Canup2002}
{Canup} R.~M.,  {Ward} W.~R.,  2002, \mn@doi [\aj] {10.1086/344684}, \href
  {https://ui.adsabs.harvard.edu/abs/2002AJ....124.3404C} {124, 3404}

\bibitem[\protect\citeauthoryear{{Christiaens}, {Cantalloube}, {Casassus},
  {Price}, {Absil}, {Pinte}, {Girard}  \& {Montesinos}}{{Christiaens}
  et~al.}{2019}]{Christiaens2019}
{Christiaens} V.,  {Cantalloube} F.,  {Casassus} S.,  {Price} D.,  {Absil} O.,
  {Pinte} C.,  {Girard} J.,   {Montesinos} M.,  2019, in AAS/Division for
  Extreme Solar Systems Abstracts. p. 101.05

\bibitem[\protect\citeauthoryear{{D'Angelo}, {Durisen}  \&
  {Lissauer}}{{D'Angelo} et~al.}{2010}]{Dangelo2010}
{D'Angelo} G.,  {Durisen} R.~H.,   {Lissauer} J.~J.,  2010, in {Seager} S.,
  ed., , Exoplanets.
pp 319--346, \mn@doi{10.48550/arXiv.1006.5486}

\bibitem[\protect\citeauthoryear{{Goldreich} \& {Tremaine}}{{Goldreich} \&
  {Tremaine}}{1980}]{Goldreich1980}
{Goldreich} P.,  {Tremaine} S.,  1980, \mn@doi [\apj] {10.1086/158356}, \href
  {https://ui.adsabs.harvard.edu/abs/1980ApJ...241..425G} {241, 425}

\bibitem[\protect\citeauthoryear{{G{\"u}nther} \& {Kley}}{{G{\"u}nther} \&
  {Kley}}{2002}]{Gunther2002}
{G{\"u}nther} R.,  {Kley} W.,  2002, \mn@doi [\aap]
  {10.1051/0004-6361:20020407}, \href
  {https://ui.adsabs.harvard.edu/abs/2002A&A...387..550G} {387, 550}

\bibitem[\protect\citeauthoryear{{Haffert}, {Bohn}, {de Boer}, {Snellen},
  {Brinchmann}, {Girard}, {Keller}  \& {Bacon}}{{Haffert}
  et~al.}{2019}]{Haffert2019}
{Haffert} S.~Y.,  {Bohn} A.~J.,  {de Boer} J.,  {Snellen} I.~A.~G.,
  {Brinchmann} J.,  {Girard} J.~H.,  {Keller} C.~U.,   {Bacon} R.,  2019,
  \mn@doi [Nature Astronomy] {10.1038/s41550-019-0780-5}, \href
  {https://ui.adsabs.harvard.edu/abs/2019NatAs...3..749H} {3, 749}

\bibitem[\protect\citeauthoryear{{Hanawa}, {Ochi}  \& {Ando}}{{Hanawa}
  et~al.}{2010}]{Hanawa2010}
{Hanawa} T.,  {Ochi} Y.,   {Ando} K.,  2010, \mn@doi [\apj]
  {10.1088/0004-637X/708/1/485}, \href
  {https://ui.adsabs.harvard.edu/abs/2010ApJ...708..485H} {708, 485}

\bibitem[\protect\citeauthoryear{{Hashimoto} et~al.,}{{Hashimoto}
  et~al.}{2012}]{Hashimoto2012}
{Hashimoto} J.,  et~al., 2012, \mn@doi [\apjl] {10.1088/2041-8205/758/1/L19},
  \href {https://ui.adsabs.harvard.edu/abs/2012ApJ...758L..19H} {758, L19}

\bibitem[\protect\citeauthoryear{{Kisare} \& {Fabrycky}}{{Kisare} \&
  {Fabrycky}}{2024}]{Kisare2024}
{Kisare} A.~M.,  {Fabrycky} D.~C.,  2024, \mn@doi [\mnras]
  {10.1093/mnras/stad3543}, \href
  {https://ui.adsabs.harvard.edu/abs/2024MNRAS.527.4371K} {527, 4371}

\bibitem[\protect\citeauthoryear{{Kley}}{{Kley}}{1999}]{Kley1999}
{Kley} W.,  1999, \mn@doi [\mnras] {10.1046/j.1365-8711.1999.02198.x}, \href
  {https://ui.adsabs.harvard.edu/abs/1999MNRAS.303..696K} {303, 696}

\bibitem[\protect\citeauthoryear{{Krapp}, {Kratter}, {Youdin},
  {Ben{\'\i}tez-Llambay}, {Masset}  \& {Armitage}}{{Krapp}
  et~al.}{2024}]{Krapp2024}
{Krapp} L.,  {Kratter} K.~M.,  {Youdin} A.~N.,  {Ben{\'\i}tez-Llambay} P.,
  {Masset} F.,   {Armitage} P.~J.,  2024, \mn@doi [arXiv e-prints]
  {10.48550/arXiv.2402.14638}, \href
  {https://ui.adsabs.harvard.edu/abs/2024arXiv240214638K} {p. arXiv:2402.14638}

\bibitem[\protect\citeauthoryear{{Larwood}, {Nelson}, {Papaloizou}  \&
  {Terquem}}{{Larwood} et~al.}{1996}]{Larwood1996}
{Larwood} J.~D.,  {Nelson} R.~P.,  {Papaloizou} J.~C.~B.,   {Terquem} C.,
  1996, MNRAS, \href {http://adsabs.harvard.edu/abs/1996MNRAS.282..597L} {282,
  597}

\bibitem[\protect\citeauthoryear{{Lissauer}, {Hubickyj}, {D'Angelo}  \&
  {Bodenheimer}}{{Lissauer} et~al.}{2009}]{Lissauer2009}
{Lissauer} J.~J.,  {Hubickyj} O.,  {D'Angelo} G.,   {Bodenheimer} P.,  2009,
  \mn@doi [\icarus] {10.1016/j.icarus.2008.10.004}, \href
  {https://ui.adsabs.harvard.edu/abs/2009Icar..199..338L} {199, 338}

\bibitem[\protect\citeauthoryear{{Lubow} \& {Ogilvie}}{{Lubow} \&
  {Ogilvie}}{2000}]{Lubow2000}
{Lubow} S.~H.,  {Ogilvie} G.~I.,  2000, \mn@doi [\apj] {10.1086/309101}, \href
  {https://ui.adsabs.harvard.edu/abs/2000ApJ...538..326L} {538, 326}

\bibitem[\protect\citeauthoryear{{Lubow}, {Seibert}  \& {Artymowicz}}{{Lubow}
  et~al.}{1999}]{Lubow1999}
{Lubow} S.~H.,  {Seibert} M.,   {Artymowicz} P.,  1999, \mn@doi [\apj]
  {10.1086/308045}, \href
  {https://ui.adsabs.harvard.edu/abs/1999ApJ...526.1001L} {526, 1001}

\bibitem[\protect\citeauthoryear{{Martin} \& {Armitage}}{{Martin} \&
  {Armitage}}{2021}]{Martin2021a}
{Martin} R.~G.,  {Armitage} P.~J.,  2021, \mn@doi [\apjl]
  {10.3847/2041-8213/abf736}, \href
  {https://ui.adsabs.harvard.edu/abs/2021ApJ...912L..16M} {912, L16}

\bibitem[\protect\citeauthoryear{{Martin} \& {Lubow}}{{Martin} \&
  {Lubow}}{2011}]{Martin2011}
{Martin} R.~G.,  {Lubow} S.~H.,  2011, \mn@doi [\mnras]
  {10.1111/j.1365-2966.2011.18228.x}, \href
  {https://ui.adsabs.harvard.edu/abs/2011MNRAS.413.1447M} {413, 1447}

\bibitem[\protect\citeauthoryear{{Martin}, {Zhu}  \& {Armitage}}{{Martin}
  et~al.}{2020}]{Martin2020}
{Martin} R.~G.,  {Zhu} Z.,   {Armitage} P.~J.,  2020, \mn@doi [\apjl]
  {10.3847/2041-8213/aba3c1}, \href
  {https://ui.adsabs.harvard.edu/abs/2020ApJ...898L..26M} {898, L26}

\bibitem[\protect\citeauthoryear{{Martin}, {Armitage}, {Lubow}  \&
  {Price}}{{Martin} et~al.}{2023}]{Martin2023}
{Martin} R.~G.,  {Armitage} P.~J.,  {Lubow} S.~H.,   {Price} D.~J.,  2023,
  \mn@doi [\apj] {10.3847/1538-4357/ace345}, \href
  {https://ui.adsabs.harvard.edu/abs/2023ApJ...953....2M} {953, 2}

\bibitem[\protect\citeauthoryear{{Morbidelli}, {Szul{\'a}gyi}, {Crida}, {Lega},
  {Bitsch}, {Tanigawa}  \& {Kanagawa}}{{Morbidelli}
  et~al.}{2014}]{Morbidelli2014}
{Morbidelli} A.,  {Szul{\'a}gyi} J.,  {Crida} A.,  {Lega} E.,  {Bitsch} B.,
  {Tanigawa} T.,   {Kanagawa} K.,  2014, \mn@doi [\icarus]
  {10.1016/j.icarus.2014.01.010}, \href
  {https://ui.adsabs.harvard.edu/abs/2014Icar..232..266M} {232, 266}

\bibitem[\protect\citeauthoryear{{Mosqueira} \& {Estrada}}{{Mosqueira} \&
  {Estrada}}{2003}]{Mosqueira2003}
{Mosqueira} I.,  {Estrada} P.~R.,  2003, \mn@doi [\icarus]
  {10.1016/S0019-1035(03)00076-9}, \href
  {https://ui.adsabs.harvard.edu/abs/2003Icar..163..198M} {163, 198}

\bibitem[\protect\citeauthoryear{{Ogilvie}}{{Ogilvie}}{2006}]{Ogilvie2006}
{Ogilvie} G.~I.,  2006, \mn@doi [\mnras] {10.1111/j.1365-2966.2005.09776.x},
  \href {https://ui.adsabs.harvard.edu/abs/2006MNRAS.365..977O} {365, 977}

\bibitem[\protect\citeauthoryear{{Papaloizou} \& {Lin}}{{Papaloizou} \&
  {Lin}}{1984}]{Papaloizou1984}
{Papaloizou} J.,  {Lin} D.~N.~C.,  1984, \mn@doi [\apj] {10.1086/162561}, \href
  {https://ui.adsabs.harvard.edu/abs/1984ApJ...285..818P} {285, 818}

\bibitem[\protect\citeauthoryear{{Papaloizou} \& {Lin}}{{Papaloizou} \&
  {Lin}}{1995}]{Papaloizou1995a}
{Papaloizou} J.~C.~B.,  {Lin} D.~N.~C.,  1995, \mn@doi [\apj] {10.1086/175127},
  \href {https://ui.adsabs.harvard.edu/abs/1995ApJ...438..841P} {438, 841}

\bibitem[\protect\citeauthoryear{{Papaloizou} \& {Terquem}}{{Papaloizou} \&
  {Terquem}}{1995}]{Papaloizou1995}
{Papaloizou} J.~C.~B.,  {Terquem} C.,  1995, MNRAS, \href
  {http://adsabs.harvard.edu/abs/1995MNRAS.274..987P} {274, 987}

\bibitem[\protect\citeauthoryear{{Peale} \& {Lee}}{{Peale} \&
  {Lee}}{2002}]{Peale2002}
{Peale} S.~J.,  {Lee} M.~H.,  2002, \mn@doi [Science]
  {10.1126/science.1076557}, \href
  {https://ui.adsabs.harvard.edu/abs/2002Sci...298..593P} {298, 593}

\bibitem[\protect\citeauthoryear{{Pollack}, {Hubickyj}, {Bodenheimer},
  {Lissauer}, {Podolak}  \& {Greenzweig}}{{Pollack} et~al.}{1996}]{Pollack1996}
{Pollack} J.~B.,  {Hubickyj} O.,  {Bodenheimer} P.,  {Lissauer} J.~J.,
  {Podolak} M.,   {Greenzweig} Y.,  1996, \mn@doi [\icarus]
  {10.1006/icar.1996.0190}, \href
  {https://ui.adsabs.harvard.edu/abs/1996Icar..124...62P} {124, 62}

\bibitem[\protect\citeauthoryear{{Sefilian} \& {Rafikov}}{{Sefilian} \&
  {Rafikov}}{2019}]{Sefilian2019}
{Sefilian} A.~A.,  {Rafikov} R.~R.,  2019, \mn@doi [\mnras]
  {10.1093/mnras/stz2412}, \href
  {https://ui.adsabs.harvard.edu/abs/2019MNRAS.489.4176S} {489, 4176}

\bibitem[\protect\citeauthoryear{{Szul{\'a}gyi}, {Masset}, {Lega}, {Crida},
  {Morbidelli}  \& {Guillot}}{{Szul{\'a}gyi} et~al.}{2016}]{Szulagyi2016}
{Szul{\'a}gyi} J.,  {Masset} F.,  {Lega} E.,  {Crida} A.,  {Morbidelli} A.,
  {Guillot} T.,  2016, \mn@doi [\mnras] {10.1093/mnras/stw1160}, \href
  {https://ui.adsabs.harvard.edu/abs/2016MNRAS.460.2853S} {460, 2853}

\bibitem[\protect\citeauthoryear{{Tremaine}}{{Tremaine}}{2023}]{Tremaine2023}
{Tremaine} S.,  2023, {Dynamics of Planetary Systems}

\bibitem[\protect\citeauthoryear{{Tremaine}, {Touma}  \& {Namouni}}{{Tremaine}
  et~al.}{2009}]{Tremaine2009}
{Tremaine} S.,  {Touma} J.,   {Namouni} F.,  2009, \mn@doi [\aj]
  {10.1088/0004-6256/137/3/3706}, \href
  {https://ui.adsabs.harvard.edu/abs/2009AJ....137.3706T} {137, 3706}

\bibitem[\protect\citeauthoryear{{Ward}}{{Ward}}{1981}]{Ward1981}
{Ward} W.~R.,  1981, \mn@doi [\icarus] {10.1016/0019-1035(81)90079-8}, \href
  {https://ui.adsabs.harvard.edu/abs/1981Icar...46...97W} {46, 97}

\bibitem[\protect\citeauthoryear{{Zanazzi} \& {Lai}}{{Zanazzi} \&
  {Lai}}{2018}]{Zanazzi2018a}
{Zanazzi} J.~J.,  {Lai} D.,  2018, \mn@doi [\mnras] {10.1093/mnras/sty951},
  \href {https://ui.adsabs.harvard.edu/abs/2018MNRAS.477.5207Z} {477, 5207}

\makeatother
\end{thebibliography}



\appendix
\section{Analytic Expression for $\beta_2$ in a Nonself-Gravitating and Inviscid Disc}

\label{sec:appbeta2}
We provide an analytic expression for $\beta_2(x)$ in Equation (\ref{betaexp}) using Equations (\ref{beta0}) to (\ref{cint}). 
The result is expressed as
\begin{equation}
\beta_2(x) = a_0 \left(a_1 + a_2 x^{-4} + a_3 x^{-2}
+ a_4 x + a_5 x^3 +a_6 x^6 \right), \label{betastdexp}
\end{equation}
for coefficients $a_j$.
We define 
\begin{align}
    & b_1 = x_{\rm in}+ x_{\rm out}, \\
    & b_2 = x_{\rm in}^2+  x_{\rm in} x_{\rm out} + x_{\rm out}^2,
\end{align}
and then have that
\begin{align}
& a_0 = \frac{\beta_{\rm s}- \beta_{\rm p}}{3 x_{\rm out} + x_{\rm in} 
( 3 + 2 x_{\rm in} x_{\rm out}^2 b_2) }, \\
 &  a_1=  \frac{1}{36 x_{\rm in}^2 x_{\rm out}^2}[21 x_{\rm in}^7 x_{\rm out}^4 + 2 x_{\rm out}^2 b_1 ( 3 x_{\rm out}^3 -2 x_{\rm in}^8) + 
  4 x_{\rm in}^2 x_{\rm out}^4 (3 - x_{\rm out}^5) +  \\ &
 \qquad  \qquad
   x_{\rm in}^3 x_{\rm out}^3 (87 - 4 x_{\rm out}^5) + 3 x_{\rm in}^4 x_{\rm out}^2 (4 + 7 x_{\rm out}^5) + 
   x_{\rm in}^6 (6 + 17 x_{\rm out}^5) + x_{\rm in}^5 x_{\rm out} (6 + 17 x_{\rm out}^5)], \\
& a_2 = \frac{x_{\rm in}^2 x_{\rm out}^2 b_2}{4}, \\
& a_3 = -\frac{x_{\rm in}^3 b_1 + x_{\rm out}^2 b_2}{2}, \\
& a_4 = -b_1 - x_{\rm in}^2 x_{\rm out}^2  b_2, \\
& a_5 = \frac{x_{\rm in}^3 b_1 + x_{\rm out}^2 b_2}{3}, \\
    & a_6 = -\frac{b_1}{6}. 
\end{align}
Boundary conditions given by Equation (\ref{BCf}) are satisfied since
 $ \dd \beta_2(x_{\rm in})/\dd x= \dd\beta_2(x_{\rm out})/\dd x=0$.

\section{Comparison of Tilt Determination Methods }
\label{sec:comp}

In Section \ref{sec:warm} we applied an expansion method in $1/p$ to determine the  disc tilt. The method is applicable to mildly warped discs that do not contain a Laplace radius because $x_{\rm in} > x_{\rm L, in}$ and $x_{\rm out} < x_{\rm L, out}$. But there are no other restrictions on $x_{\rm in}$ and $x_{\rm out}$. In Section \ref{sec:cool} we applied a boundary layer approach to determine the disc tilt that is not restricted by this inner and radius condition. It assumes that there is a large change in scale from the inner to outer regions of the disc, $x_{\rm in} \ll 1 \ll x_{\rm out}$. The tilt is determined at small $x$ by accounting for the planetary torque and ignoring the stellar torque that dominates at large $x$ and vice versa at large $x$. The tilt at intermediate radii is determined by the constraint that the inner and outer tilt and flux match.

 We compare the results obtained from the two methods. To make this comparison, we retain only the lowest order term in $x_{\rm out}/x_{\rm in}$ to Equation (\ref{beta0}) for $\beta_0$ and for each coefficient $a_j$ in Equation (\ref{betastdexp})  for $\beta_2$. We also retain terms involving $x_{\rm in}^2 x_{\rm out}^3$. Equation (\ref{beta0})  reduces to
 Equation (\ref{beta0xixo}).
 Equation  (\ref{betastdexp})  reduces to \begin{equation} \beta_2(x)= \frac{\beta_{\rm s}- \beta_{\rm p}}  { 36  (3  + 2 x_{\rm in}^2 x_{\rm out}^3)} \left[ x_{\rm out}^3 \left(\frac{6}{x_{\rm in}^2} - 4 x_{\rm out}^3 \right) + \frac{9 x_{\rm in}^2 x_{\rm out}^3}{ x^4}   -\frac{18 x_{\rm out}^3}{x^2}  -36 x (1 + x_{\rm in}^2 x_{\rm out}^3)  + 12 x^3 x_{\rm out}^3  -6 x^6 \right]. \label{b2xixoexp} \end{equation}
 This simplified expression for $\beta(x)$
 provides nearly identical values as the full expression given by Equation (\ref{betastdexp}) for the warm disc model
 plotted in Figure~\ref{fig:betastd}.

To compare with the boundary layer results in Section \ref{sec:cool}, we first expand the inner solution for $\beta$
given by Equation (\ref{betain}) to order $1/p^2$.
The result to lowest order agrees with Equation (\ref{beta0xixo}) for $\beta_0$. At order $1/p^2$, the results match
those of Equation (\ref{b2xixoexp}) for $\beta_2$ for the constant term,
 the $1/x^2$ term and the $1/x^4$ term.

We next expand the outer solution for $\beta$
given by Equation (\ref{betaout}) to order $1/p^2$. Again,
the result to lowest order agrees with Equation (\ref{beta0xixo}). At order $1/p^2$, the results match
those of Equation (\ref{b2xixoexp}) for the constant term,
the $x^3$ term and the $x^6$ term. The $x$ term 
in (\ref{b2xixoexp}) is not present in the expansions of either the inner or outer solutions.  However, for $x_{\rm in} \ll 1  \ll  x_{\rm out}$, this term is  small compared to sum of the contributions from the other terms that vary in $x$.

\section{Analytic determination of $\beta$ for a Viscous Non-self-gravitating Disc}
\label{sec:betaviscanalytic}

We solve Equations (\ref{betavisc}) and (\ref{betavisc}) with boundary conditions given by Equation
(\ref{BCf}) by means of a perturbation expansion in $\epsilon = 1/p$. We extend the approach
taken in Section \ref{sec:warm} to include non-zero viscosity.

We have that
\begin{equation}
\epsilon \left[ \frac{\beta - \beta_{\rm p}}{x^3} + (\beta-\beta_{\rm s}) x^2  \right]= \frac{\dd f}{\dd x}, \label{betavisca}
\end{equation}
and
\begin{equation}
\epsilon \left(\frac{1}{x^3} + x^2 \right) f  + \frac{2 \ii \alpha  f }{h x}=  \frac{\dd \beta}{\dd x} \label{fvisca},
\end{equation}
We apply expansions
\begin{equation}
\beta =  \beta_0 + \epsilon \beta_1 + \cdots
 \end{equation}
 and \begin{equation}
 f =  f_0 + \epsilon f_1 + \cdots,
 \end{equation}
regarding $\alpha$, $h$, $x$, $f$, and $\beta$ as order unity quantities in the expansion in $\epsilon$.

To lowest order,
 we again recover $\beta_0$ and $f_1(x)$ given by Equations (\ref{beta0}) and (\ref{f1}), respectively.
In next order in Equation (\ref{fvisca}) we have that
\begin{equation}
\beta_1(x) =  \frac{2 \ii \alpha}{h} \int \frac{f_1(x)}{x} \, \dd x + c_1,
\end{equation}
where $c_1$ is a constant of integration to be determined.
Equation (\ref{betavisca}) implies
 \begin{equation}
 f_2(x) =  \int_{x_{\rm in}}^x \left(\frac{1}{x'^3}+x'^2 \right) \beta_1(x') \, \dd x'.
 \end{equation}
We solve for $c_1$ by requiring that $f_2(x_{\rm out}) =0$.
We then determine $\beta_1(x)$ and $f_2(x)$.
In next order in Equation (\ref{fvisca}),
 \begin{equation}
 \beta_2(x) =  \int \left [\left(\frac{1}{x^3}+x^2 \right) f_1(x) +\frac{2 \ii \alpha  f_2(x) }{h x} \right] \, \dd x  + c_2,
 \end{equation}
 where $c_2$ is a constant of integration to be determined.
 In next order in Equation (\ref{betavisca}),
 we have an integral constraint
 \begin{equation}
 \int_{x_{\rm in}}^{x_{\rm out}} \left ( \frac{1}{x^3} +  x^2 \right)   \beta_2(x) \, \dd x =0. \label{c2int}
 \end{equation}
 This constraint determines $c_2$ which completes the solution for $\beta_2(x)$.

We then calculate
\begin{equation}
    \beta(x) \simeq \beta_0 + \epsilon \beta_1(x) + \epsilon^2 \beta_2(x). \label{betaviscexp} 
\end{equation}
Constant tilt $\beta_0$ is given by Equation (\ref{beta0}).  $\beta_1(x)$ 
is purely imaginary, while $\beta_2(x)$ is purely real. The terms that
depend on $\alpha$ are denoted by $\delta \beta$.
The expression for $\delta \beta$ is quite long. To provide approximate simpler  expressions,
 we consider the case that $x_{\rm in} \ll 1 \ll x_{\rm out}$ such that  
 $x_{\rm in}^2 x_{\rm out}^3 \sim 1$.
 These conditions are satisfied for the models we consider in this paper.
 We find in lowest order of $x_{\rm in}$ that
\begin{equation}
 \epsilon \, \delta \beta_1(x) = \ii (\beta_{\rm s} - \beta_{\rm p}) \left(\frac{\alpha}{h} \right)
  \left( \frac{x_{\rm out}^{3/2}}{p  x_{\rm in}}  \right)
 \frac{k_1 + k_2 \log(x)}{k_3}, \label{b1visc}
\end{equation}
with
\begin{align}
k_1 = \xi^{1/2} (-9 -12 \log(x_{\rm in}) +4 \xi( 1- 2 \log(x_{\rm out}))),
\end{align}
\begin{align}
k_2 =  4 \xi^{1/2} (3+ 2 \xi),
\end{align}
and
\begin{align}
k_3=  2 (3 + 2 \xi)^2,
\end{align}
where $\xi= x_{\rm in}^2 x_{\rm out}^3$ and $\log$ denotes the natural logarithm. Also
\begin{align}
\epsilon^2 \delta \beta_2(x) &=  (\beta_{\rm s} - \beta_{\rm p}) \left(\frac{\alpha}{h} \right)^2
  \left( \frac{x_{\rm out}^3}{p^2  x_{\rm in}^2}  \right) \frac{k_4 + k_5 \log(x)}{k_6}, \label{b2visc}
\end{align}
where
\begin{align}
k_4 = &[(9 - 4 \xi) (189 + 1020 \xi + 56 \xi^2) +
 648 \xi (12 \log(x_{\rm in})^2 - 9 \log(x_{\rm out}) + \\
    & 2 \xi \log(x_{\rm out}) (7 - 4 \log(x_{\rm out})) + 
    4 \log(x_{\rm in}) (6 - \xi + (-3 + 2 \xi) \log(x_{\rm out})))]/648,
\end{align}
\begin{align}
k_5 =  -\xi (3 + 2 \xi) (5 + 4 \log(x_{\rm in}/x_{\rm out})),
\end{align}
and
\begin{align}
k_6=  (3 + 2 \xi)^3.
\end{align}

\begin{figure}
    \includegraphics[width=0.8\textwidth]{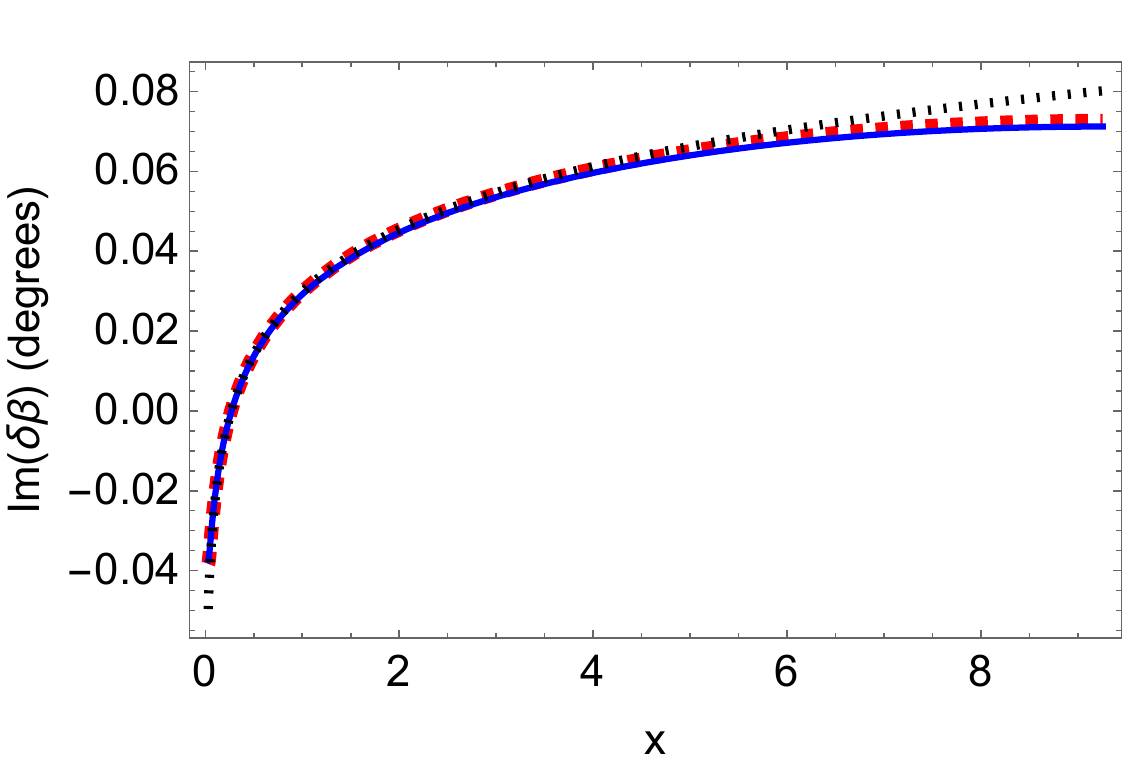}
    \caption{Plot of  $\text{Im}(\delta \beta(x))$ for $\alpha=0.005$ for Model F. The solid blue line is from the numerical solution, the red dashed line is from the full series solution for $\text{Im}(\epsilon  \delta \beta_1)$, and the dotted black line is from the approximate series solution for $\text{Im}(\epsilon \delta \beta_1)$  given by the imaginary part of Equation (\ref{b1visc}).}
    \label{fig:betasImapprox}
\end{figure}

The terms in parentheses in Equations (\ref{b1visc}) and (\ref{b2visc}) that contain $p$ are somewhat
less than unity for Model F, since $x_{\rm in} > 1/\sqrt{p}  =x_{\rm L, in}$ and $x_{\rm out} < p^{1/3} =x_{\rm L, out}$, as discussed in Sections \ref{sec:warm} and  \ref{sec:cool}. 
For $\alpha =0.005$ in Model F, the error in the approximate expressions
given by Equations (\ref{b1visc}) and (\ref{b2visc}) is about 10\%.
Figure \ref{fig:betasImapprox} plots the 
imaginary part of $\delta \beta(x)$ in that case. 

\bsp	
\label{lastpage}
\end{document}